\documentclass[10pt]{article}

\usepackage[T1]{fontenc}
\usepackage[utf8]{inputenc}
\usepackage[english]{babel}
\usepackage{amssymb,amsmath,latexsym,amsfonts,amsthm,bm,mathtools}
\usepackage{calrsfs}
\usepackage{graphicx}
\usepackage[top=1.5in, bottom=1.5in, left=1.25in, right=1.25in]{geometry}
\usepackage{booktabs,url,tikz,multirow,bigints}

\title{A state space approach to dynamic modeling of mouse-tracking data}
\author{Antonio Calcagn\`{i}\footnote{Corresponding author: antonio.calcagni@unipd.it}, Luigi Lombardi$^\dagger$, Marco D'Alessandro$^\dagger$\\ \\
 $^*$University of Padova\\
 $^\dagger$University of Trento
}

\date{}

\DeclareMathAlphabet{\mathcall}{OMS}{zplm}{m}{n}

\begin{document}

\maketitle

\begin{abstract}
Mouse-tracking recording techniques are becoming very attractive in experimental psychology. They provide an effective means of enhancing the measurement of some real-time cognitive processes involved in categorization, decision-making, and lexical decision tasks. Mouse-tracking data are commonly analysed using a two-step procedure which first summarizes individuals' hand trajectories with independent measures, and then applies standard statistical models on them. However, this approach can be problematic in many cases. In particular, it does not provide a direct way to capitalize the richness of hand movement variability within a consistent and unified representation. In this article we present a novel, unified framework for mouse-tracking data. Unlike standard approaches to mouse-tracking, our proposal uses stochastic state-space modeling to represent the observed trajectories in terms of both individual movement dynamics and experimental variables. The model is estimated via a Metropolis-Hastings algorithm coupled with a non-linear recursive filter. The characteristics and potentials of the proposed approach are illustrated using a lexical decision case study. The results highlighted how dynamic modeling of mouse-tracking data can considerably improve the analysis of mouse-tracking tasks and the conclusions researchers can draw from them.  \\\vspace{0.15cm}

\end{abstract}

\section{Introduction}

Over the last decades, the study of computer-mouse trajectories has brought to light new perspectives into the investigation of a wide range of cognitive processes (e.g., for a recent review see \cite{freeman2017}). Unlike traditional behavioral measures such as reaction times and accuracies, mouse trajectories may offer a valid and cost-effective way to measure the real-time evolution of ongoing cognitive processes during experimental tasks \cite{friedman2013linking}. This has also been supported by recent researches investigating mouse-tracking in association to more consolidated experimental devices, such as eye-tracking and fMRI \cite{quetard2016differential,stolier2017neural}. In a typical mouse-tracking experiment, participants are presented with a computer-based interface showing the stimulus at the bottom of the screen and two competing categories on the left and right top corners. Participants are asked to select the most appropriate label given the task instruction and stimulus while the x-y trajectories are instantaneously recorded. The main idea is that trajectories of reaching movements can unfold the decision process underlying the hand movement behavior. For instance, the curvature of computer-mouse trajectories might reveal competing processes activated in discriminating the two categories. Mouse-tracking has been successfully applied in several cognitive research studies, including lexical decision \cite{ke2017quantificational,incera2017sentence}, social categorization \cite{freeman2016perceptual,carraro2016hand}, numerical cognition \cite{faulkenberry2014hand,faulkenberry2016testing}, memory \cite{papesh2012memory}, moral decision \cite{koop2013assessment}, and lie detection \cite{monaro2017detection}. Moreover, the availability of specialized and freely-available software for mouse-tracking experiments have strongly contributed to the wide-spread application of such a methodology in the more general psychological domain \cite{freeman2010mousetracker,kieslich2017mousetrap}. Recently, the debate on the nature of cognitive processes tracked by this type of reaching trajectories have also received attention from the motor control literature \cite{van2009trajectories,friedman2013linking}. 

So far, mouse-tracking data have been analysed using simple strategies based on the conversion of x-y trajectories into summary measures, such as maximum deviation, area under the curve, response time, initiation time \cite{hehman2014advanced}. Although these steps are still meaningful in case of simple and well-behaved x-y trajectories, they can also provide biased results if applied to more complex and possibly noisy data. To circumvent these problems, other approaches have been proposed more recently \cite{cox2012gaussian,calcagni2017analyzing,zgonnikov2017decision,krpan2017behavioral}. However, also the more recent proposals require modeling empirical trajectories before the data-analysis. Although these approaches potentially provide informative results in many empirical cases, they can also suffer from a number of issues, which revolve around the reduction of x-y data to simple scalar measures. For instance, problems may arise in the case of trajectories showing multiple phases, averaging with non-homogeneous curves, and signal-noise discrimination \cite{calcagni2017analyzing}. As far as we know, a proper framework to simultaneously model and analyse mouse-tracking data in a unified way is still lacking. 

In this paper we describe an alternative perspective based on a state-space approach with the aim to simultaneously model and analyse mouse-tracking data. State-space models are very general time-series methods that allow estimating unobserved dynamics which gradually evolve over discrete time. As for diffusion models, which are widely used in modeling the temporal evolution of cognitive decision processes \cite{smith2004psychology}, they belong to the general family of stochastic processes and offer optimal discrete approximation to many continuous differential systems used to represent dynamics with autoregressive patterns \cite{cox2017theory}. In particular, we used a non-linear and discrete-time model that represents mouse trajectories as a function of some typical experimental manipulations. The model is estimated under a Bayesian framework, using a conjunction of a non-linear recursive filter and a Metropolis-Hastings algorithm. Data analyses is then performed using posterior distributions of model parameters \cite{gelman2014bayesian}. 

The reminder of this article is organized as follows. In Section 2 we motivate our proposal by reviewing the main issues of a typical mouse-tracking experiment. In Section 3 we present our proposal and describe its main characteristics. In Section 4 we describe the application of our method to a psycholinguistic case study. Section 5 provides a general discussion of the results, comments and suggestions for further investigations. Section 6 concludes the article by summarizing its main findings. 	

\section{A motivating example}
	
\begin{figure}[!h]
	\centering
	\includegraphics[scale=0.7]{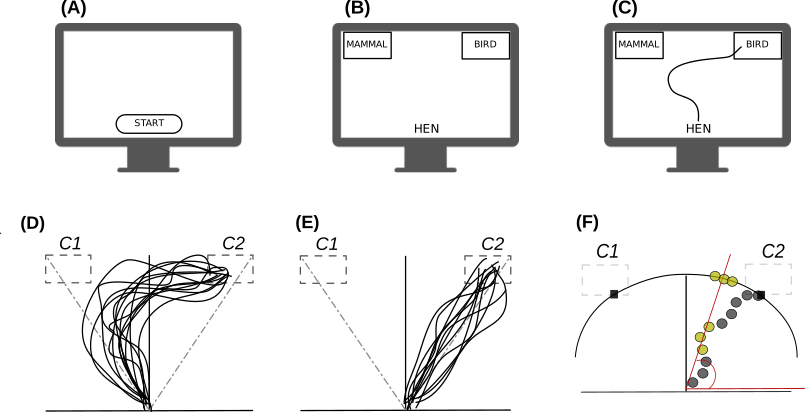}
	\caption{(A)-(C) Conceptual diagram of a typical mouse-tracking task: (A)-(B) stimulus presentation, (B) participant's response. (D)-(E) Prototypical mouse-tracking trajectories collapsed over participants and trials as a function of manipulation task: (D) case where the manipulation has an effect - as revealed by the curvature of the trajectories, (E) case where the manipulation has no effect. (F) Conceptual diagram for the atan2 conversion: grey circles represent the sampled x-y trajectories, yellow circles represent those x-y pairs projected onto the circumference outer the Cartesian plane, whereas red lines represent the projection direction. Note that in a two-choice categorization task, the correct category C2 is presented on the top-right label (\textit{target}) whereas the competing category C1 is presented in the opposite top-left label (\textit{distractor}). }
	\label{fig1}
\end{figure}

To begin with, consider a two-choice semantic categorization task \cite{dale2007graded}, in which participants have to classify semantic stimuli (e.g., name of animals) into their corresponding categories (e.g., mammal, fish). In the most typical implementation of a mouse-tracking task, participants would sit in front of a computer screen showing a resting frame (see Figure \ref{fig1}-a). They start a trial by clicking a starting button at the bottom-center of the screen, after which they are presented with a given stimulus (e.g. hen, see Figure \ref{fig1}-b). To finalize the trial, participants move the cursor on the screen by means of a well tuned computer-mouse in order to reach and select one of the two labels presented on the top-left and top-right corners of the screen (e.g., mammal vs. bird, see Figure \ref{fig1}-c). In the meanwhile, x-y coordinates, initiation time, and final clicking time are registered for each participant and trial. The basic idea is that x-y trajectories reflect the extent to which the real-time categorization response is affected by the experimental manipulation. More precisely, as a result of the assumption that co-activation of competing processes continuously drive the explicit hand response \cite{spivey2006continuous}, one would suppose to see more curved - or generally irregular - trajectories in association with stimuli showing higher ambiguity. In our case, for instance, it would be expected that atypical exemplars, such as hen, dolphin, and penguin, globally produce more curved or irregular trajectories than typical exemplars like dog, rabbit, and lion (see Figure \ref{fig1}-d/e). 

In the mouse-tracking literature, data analysis commonly proceeds summarizing the recorded trajectories by means of few indices, which are then used as input to standard statistical techniques. In the current example, for instance, the typicality manipulation could be tested by assessing whether the distribution of maximum deviations (i.e., the maximum curvature showed by trajectories) over trials and participants is bimodal or not \cite{freeman2013assessing}. In a similar way, linear models could be employed to test whether the typicality effect varies as a function of external covariates, such as psycholinguistic variables. 

However, the two-step approach does have some issues. For instance, it lacks a way to represent both the experimental variability - that is induced by task manipulations - and individual variability - that is instead produced by individual-specific motor programs. Likewise, in some cases, it might neglect relevant characteristics of x-y data, with the consequence that similar classes of trajectories are treated as if they were different. Still, a two-step approach does ignore the data generation process underlying observed trajectories. This does not allow, for example, making predictions or simulations on new data given the experimental settings. 

In the next section, we will present a dynamic probabilistic model that handles mouse-tracking data in a unified way. Our proposal is based upon a Bayesian non-linear state space approach, which offers a good compromise between model flexibility and model simplicity while overcoming many drawbacks of the standard mouse-tracking analyses.  	

However, the two-step approach does have some issues. For instance, it lacks a way to represent both the experimental variability - that is induced by task manipulations - and individual variability - that is instead produced by individual-specific motor programs. Likewise, in some cases, it might neglect relevant characteristics of x-y data, with the consequence that similar classes of trajectories are treated as if they were different. Still, a two-step approach does ignore the data generation process underlying observed trajectories. This does not allow, for example, making predictions or simulations on new data given the experimental settings. 

In the next section, we will present a dynamic probabilistic model that handles mouse-tracking data in a unified way. Our proposal is based upon a Bayesian non-linear state space approach, which offers a good compromise between model flexibility and model simplicity while overcoming many drawbacks of the standard mouse-tracking analyses.

\section{State-space modeling of mouse-tracking data}

A state-space model is a mathematical description used to represent linear or generally non-linear dynamic models. In their general form, state-space systems consist of (i) a measurement density $ f_y(\mathbf y_n;\mathbf{z}_n,\boldsymbol{\theta}_y)$ that describes how the observed vector of data $\mathbf y_n$ at time step $n$ is linked to a possibly underlying process $\mathbf z_n$ and (ii) a state density $f_z(\mathbf z_n;\boldsymbol{\theta}_z)$ describing the transition dynamics that drive the vector of states $\mathbf{z}_n$. Temporal dynamics can be discrete or continuous and, in the latter case, stochastic differential equations are used to model the transition dynamics. By and large, there are two aims of any analysis involving state-space models. The first is to infer the unobserved process $\widetilde{\mathbf{Z}} = (\mathbf{z}_0,\ldots,\mathbf{z}_N)$ given the data $\mathbf Y = (\mathbf{y}_0,\ldots,\mathbf{y}_N)$. This task is usually accomplished by means of filtering and smoothing procedures \cite{jazwinski2007stochastic}. The second aim regards estimating the parameters $(\boldsymbol{\theta}_y, \boldsymbol{\theta}_z)$ given the complete set of data $(\widetilde{\mathbf{Z}},\mathbf{Y})$. This is commonly performed using gradient-based methods on the likelihood of the model \cite{shumway1982approach}. Although state space models were originally used in the area of aerospace modeling \cite{kalman1960new}, they are now applied in a wide variety of domains, including control theory, remote sensing, economics, and statistics \cite{hamilton1994state,shumway2006time}. Recently, there has also been an increasing interest in psychology, where state-space models have been used to analyse, for example, dyadic interactions \cite{song2009state}, affective dynamics \cite{lodewyckx2011hierarchical,bringmann2017changing}, facial electromyography data \cite{yang2010using}, individual differences \cite{hamaker2012regime,chow2013nonlinear}, and path analysis \cite{gu2014state}. 

In line with this, we developed a state-space representation to simultaneously model and analyse mouse-tracking data. In particular, our proposal is to represent the empirical collection of computer-mouse trajectories as a function of two independent sub-models, one representing the experimental manipulations (\textit{stimuli equation}) and the other capturing the main features of the mouse movement process (\textit{states equation}). Thus, the goal of our analysis is twofold: (i) to determine the states equation for each participant over a set of experimental trials, (ii) to estimate the parameters governing the stimuli equation. The first goal will provide information on how participants differ from each other in terms of movement dynamics. By contrast, the second goal will find out to what extent the experimental manipulations affect the individual variations in producing mouse-tracking responses. 

\subsection{Data}

Let $\mathbf S$ be a $I$ (individuals) $\times$ $J$ (trials) array representing the observed data. The element $\mathbf s_{ij}$ of $\mathbf S$ defines the sub-array containing the streaming of Cartesian coordinates of the computer mouse movements: 
\begin{equation*}
\mathbf s_{ij} = \big((\tilde{x}_{0},\tilde{y}_{0}),\ldots,(\tilde{x}_n,\tilde y_n),\ldots,(\tilde x_{N_{ij}},\tilde y_{N_{ij}})\big)
\end{equation*}
with $0$ and $N_{ij}$ being the first and the last coordinates for the $i$-th participant in the $j$-th trial. The coordinates in $\mathbf s_{ij}$ are temporally ordered $(0 < \ldots < n < \ldots < N_{ij})$ because they are usually collected while the computer-mouse is moving along its surface with a constant sampling rate. Further, to make the observed data comparable, we rescale and normalize $\mathbf s_{ij}$ as a function of a common ordered scale, which indicates the cumulative amount of progressive time from $0\%$ to $N=100\%$ \cite{tanawongsuwan2001gait,ramsay2006functional}. {Thus, the final trajectories $\mathbf s_{ij}$ lie on the real plane defined by the hyper-rectangles $[-1,1]^N \times [0,1]^N$, with the first movement being equal to $(\tilde x_{0i},\tilde y_{0i}) = (0,0)$ by convention. Since we are interested in studying the co-activation of competing processes as reflected in some spatial properties of the response - such as \textit{location}, \textit{direction}, and \textit{amplitude} of the action dynamics \cite{spivey2006continuous,freeman2017} - we need to simplify the original data structure so that these properties can easily emerge. Inspired by some of the work on this problem \cite{gowayyed2013histogram,kapsouras2014action,calcagni2017analyzing}, we reduce the dimensionality of the data by projecting $\mathbf s_{ij}$ in a proper lower-dimensional subspace of movement via the restricted four-quadrant inverse tangent mapping (atan2, see \cite{burger2010principles}) from the real coordinates to the interval $[0,\pi]^N$ as follows:} 
\begin{align*}
\underbrace{(y_{0}, \ldots,y_n,\ldots,y_N)}_{\mathbf y_{ij}} &=  \text{atan2}\big( \underbrace{(\tilde{x}_{0},\tilde{y}_{0}),\ldots,(\tilde{x}_n,\tilde y_n),\ldots,(\tilde x_{N_{ij}},\tilde y_{N_{ij}})}_{\mathbf s_{ij}} \big)
\end{align*}
\noindent where $y_{0}$ is the angle at the beginning of the process whereas $y_N$ is the angle at the end of the process. Figure \ref{fig1}-F shows a graphical example of the atan2 function for a hypothetical movement path. Finally, the array of angles $\mathbf y_{ij}$ is the input for our state-space model.

\subsection{Model representation}

	The unobserved \textit{states equation} of the model is a AR(1) Gaussian model $Z_{i,n}|Z_{i,n-1}$ with transition density equal to:
\begin{equation}\label{eq2_1}
f(z_{i,n}|z_{i,n-1},\theta) = \big(\sigma^2_i\sqrt{2\pi}\big)^{-1} \cdot \exp\bigl(-(z_{i,n}-z_{i,n-1})^2/2\sigma_i^2\bigl)
\end{equation}
which models how the movement process of the $i$-th subject changes from the step $n-1$ to the next step $n$. The stochastic dynamics for the $i$-th subject is constrained by the variance parameter $\sigma^2_i \in \mathbb{R}^+$ that represents the uncertainty about the future location $z_{i,n+1}$ given the current state $z_{i,n}$. 

The \textit{measurement equation} for the observations $\mathbf y_{ij} = (y_{0},\ldots,y_n,\ldots,y_N)$ is modeled by means of a two-component von-Mises mixture distribution with density equal to:
\begin{equation}\label{eq2_2}
f(y_{ijn}|\pi_{ijn},\theta) = f(y_{ijn}|\mu_1,\kappa_1)\pi_{ijn}  + f(y_{ijn}|\mu_2,\kappa_2)(1-\pi_{ijn})
\end{equation} 
where the generic density is the standard von-Mises law:
\begin{equation*}
f(y_{ijn}|\mu,\kappa) = \frac{1}{2\pi I_0(\kappa)}\exp\biggl(\cos(y_{ijn}-\mu)^{\kappa}\biggr)
\end{equation*}
In the density formula, the term $I_0(.) ~\triangleq~ (2\pi)^{-1} \int_0^{2\pi} e^{\kappa \cos x} dx~$ is the exponentially scaled Bessel function of order zero \cite{abramowitz1972handbook}. The parameters of the mixture density are mapped to the experimental interface of the two-choice categorization task (see Figure \ref{fig1}-d/e). In particular, the means $\{\mu_1,\mu_2\} \in [-3.14,3.14)^{2}$ are mapped to the label categories \textit{C1} and \textit{C2} whereas the concentrations $\{\kappa_1,\kappa_2\} \in \mathbb{R}^{2}_+$ indicate how the observations are spread around the means. In this sense, since $\{\mu_1,\mu_2\}$ are determined by the fixed and known positions of the labels \textit{C1} and \textit{C2} on the screen, they are not treated as parameters to be estimated. Finally, the terms ${\pi}_{ijn}$ and $(1-{\pi}_{ijn})$ are the probabilities to activate the first and second density of the von-Mises components and are expressed as function of the latent states $\mathbf{z}_{i,0:N}$ and some additional covariates. The model is Markovian, in the sense that the unobserved states $\{Z_{n};n>1\}$ form a Markov sequence and the measurements $\{Y_{n};n>1\}$ are conditionally independent given the unobserved states.

\begin{figure}[!h]
	\centering
	\includegraphics[scale=0.5]{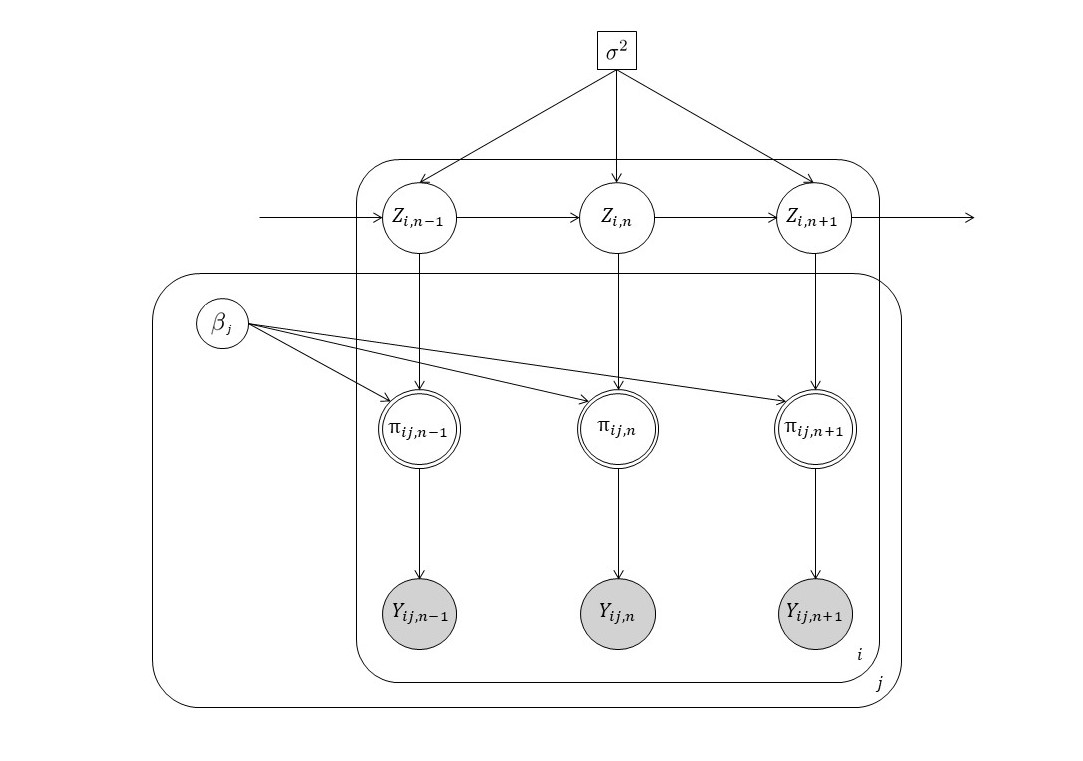}
	\caption{Graphical representation of our state-space model. Note that white circles represent unobserved random variables, white double circles indicate transformed random variables, gray circles are observed random variables. Finally, square objects depict scalar quantities. Loop over individuals $i$, trials $j$, and time steps $n$ are represented by outer squares.}
	\label{fig1mod}
\end{figure}

\begin{figure}[!h]
	\hspace{-2.5cm}
	\includegraphics[scale=0.4]{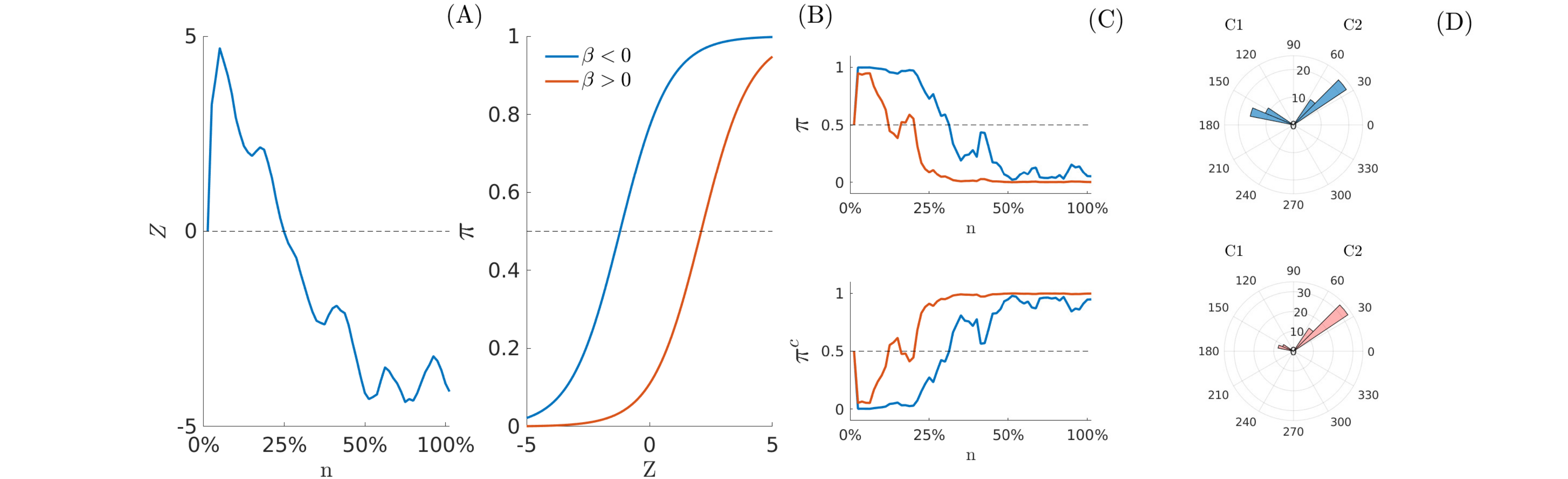}
	\caption{Conceptual diagram of the state-space representation for two hypothetical sequences of mouse-trajectories. (A) Latent movement process $\mathbf z_{0:N}$. (B) Logistic curves $\pi$ for two cases of $\boldsymbol{\beta}_J$. (C) Probability to activate the cue \textit{C1} $\boldsymbol{\pi}_{0:N}$ (upper panel) and probability to activate the cue \textit{C2} $\boldsymbol{\pi}^c_{0:N}=\boldsymbol{1}-\boldsymbol{\pi}_{0:N}$ (lower panel) for both $\beta<0$ and $\beta>0$ cases. (D) Measurements $\mathbf y_{0:N}$ as a function of their frequency (rose diagram): A case of higher attraction (upper panel) and a case of lower attraction (lower panel).}
	\label{fig1mod2}
\end{figure}

To further characterize our state-space representation, the probability $\pi_{ijn}$ is defined according to a logistic function:
\begin{equation}\label{eq1_2}
\pi_{ijn} ~\triangleq~ \big(1+\exp(-\beta_j - z_{i,n})\big)^{-1}
\end{equation}
with $\beta_j \in \mathbb{R}$ being the intercept of the model. Equation \eqref{eq1_2} can be interpreted as the probability for the $i$-th subject at step $n$ to categorize the $j$-th stimuli as belonging to \textit{C1} ($\pi_{ijn}$ tends to $1$) or \textit{C2} ($\pi_{ijn}$ tends to $0$). In addition, when the categories \textit{C1} and \textit{C2} are expressed in terms of {distractor} and {target} \cite{freeman2017}, the sequences $\boldsymbol{\pi}_{ij,0:N}$ can be interpreted as the \textit{attraction probability} that the distractor has exerted on the trajectory process $\mathbf{z}_{i,0:N}$. \\
The state-space representation is completed by linearly expanding the intercept term $\beta_j$ as follows:
\begin{equation}\label{eq1_4}
\beta_j ~\triangleq~ \sum_{k=1}^K d_{jk}\gamma_k + x_j\bigg(\eta + \sum_{k=1}^K d_{jk}\delta_k\bigg)
\end{equation}
where $\{\gamma_k,\eta,\delta_k\} \in \mathbb{R}^3$, $x_j$ is an element of the array $\mathbf x \in \mathbb{R}^J$, whereas $d_{jk} $ is an element of the (Boolean) partition matrix $\mathbf{D}_{J\times K}$, with $d_{jk} = 1$ indicating whether the $j$-th stimulus belongs to the $k$-th level of the variable $D$. Note that the matrix $\mathbf D$ satisfies the property $\sum_{k=1}^K d_{jk} = 1$, for all $j=1,\ldots,J$. In our model representation, Equation \eqref{eq1_4} is the \textit{stimuli equation} and conveys information about the experiment. It consists of three main terms. (i) A categorical term $\sum_{k=1}^K d_{jk}\gamma_k$ describing how the stimuli $\mathcall{J} = \{1,\ldots,j,\ldots,J\}$ have been arranged into $K<J$ distinct levels of a categorical variable $D$. (ii) A continuous term $x_j\eta$ that expresses the stimuli as a function of a continuous variable $X$ weighted by the coefficient $\eta$. (iii) An interaction term $x_j(\eta + \sum_{k=1}^K d_{jk} \delta_k)$ between the levels of $D$ and $X$, where $\delta_k \in \mathbb R$ and $delta_1 = 0$. This definition allows for modeling all the cases implied by an univariate experimental design with at most one covariate variable. Indeed, for $\eta = 0$ and $\boldsymbol\delta_{K} = \mathbf 0_K$ this formulation boils down to the simplest experimental case with a single categorical variable $D$. By contrast, for $\boldsymbol\delta_K = \mathbf 0_K$ and $\boldsymbol\gamma_{K} = \mathbf 0_K$ it reduces to the case where stimuli are simply paired with a continuous predictor $X$. Finally, when $\mathbf{D}_{J\times K} = \mathbf{I}_{J\times J}$, the stimuli equation reduces to the most simple case where we have as many parameters as trials. Figure \ref{fig1mod} shows a graphical representation of state-space model whereas Figure \ref{fig1mod2} illustrates the inner-working of the model for the simplest design with a two-level experimental factor.

In our model representation, the observed movement angles $\mathbf{y}_{ij,0:N}$ are sampled from \textit{C1} (resp. \textit{C2}) with probabilities equal to $\boldsymbol{\pi}_{ij,0:N}$ (resp. $\boldsymbol{\pi}^c_{ij,0:N} = 1 - \boldsymbol{\pi}_{ij,0:N}$), which in turn are expressed as a function of the AR(1) latent trajectory $\mathbf{z}_{i,0:N}$. Hence, an increase in the latent process $z_{i,n} > 0$ will also increase the probability that $y_{ijn}$ is sampled from the hemispace \textit{C1} (e.g., $\pi_{ijn} > 0.5$). By contrast, a decrease in the latent process $z_{i,n} < 0$ will increase the chance to sample $y_{ijn}$ from \textit{C2} (e.g., $\pi_{ijn} < 0.5$). As a result of Eq. \eqref{eq1_4} such an increasing (or decreasing) pattern can be modulated by the stimuli component $\boldsymbol{\beta}_J$. Moreover, as the coefficients $\boldsymbol{\beta}_J$ are decomposed as a function of $\eta$, $\boldsymbol{\gamma}_K$, and $\boldsymbol{\delta}_K$, we can also analyse the effect of $\boldsymbol{\beta}_J$ on $\boldsymbol \pi_{ij,0:N}$ in terms of the experimental manipulation $D$, the covariate $X$, or the interaction term $DX$. Figure \ref{fig1mod2} shows a conceptual representation of the modeling steps involved by our approach. Panel (A) shows an example of the random-walk used to represent the movement process (Eq. \ref{eq1_2}). Instead, panel (B) shows the logistic function used to form the \textit{stimuli equation} (Eq. \ref{eq1_2}) for two typical cases of $\boldsymbol{\beta}_J$. Panel (C) represents the probability $\boldsymbol{\pi}_{ij,0:N}$ to activate the distractor \textit{C1} (upper panel) and the probability $\boldsymbol{\pi}^c_{ij,0:N}$ to activate the target \textit{C2} (lower panel) as a function of $\mathbf{z}_{i,0:N}$ and $\boldsymbol{\beta}_J$. Finally, panel (D) depicts two cases of observed radians that are associated to $\boldsymbol{\pi}_{ij,0:N}$ and $\boldsymbol{\pi}^c_{ij,0:N}$. In particular, the upper panel shows an example of data with a pronounced attraction toward \textit{C1}, which is in turn reflected by the blue probability curve of the panels (B)-(C). By contrast, the lower panel represents data with little attraction toward \textit{C1}, as also reflected by the red probability curve of the panels (B)-(C). In this sense, as Equation \eqref{eq1_2} represents an intercept model, the parameter $\boldsymbol{\beta}_J$ does not affect the shape of the movement dynamics $\mathbf{z}_{i,0:N}$. On the contrary, it acts by shifting the movement dynamics upward ($\boldsymbol{\beta}<0$) or downward ($\boldsymbol{\beta}>0$) toward the \textit{C1} or \textit{C2} hemispaces, respectively.

\subsection{Model identification}

State-space model identification consists of inferring the unobserved sequence of states by means of filtering and smoothing algorithms and estimating the model's parameters via Likelihood-based approximations \cite{shumway2006time,sarkka2013bayesian}. For instance, in the simplest linear gaussian case, where both the states and measurement equations are linear with additive gaussian noise, inference of latent states is usually performed via Kalman filter whereas parameter's estimation is realised with the Expectation-Maximization algorithm. In our case, as Eqs.\eqref{eq1_2}-\eqref{eq1_4} describe a more complex non-linear model, we adopted a recursive \textit{Gaussian approximation filter} for the inference problem \cite{smith2003estimating}, coupled with a \textit{marginal Metropolis-Hastings} MCMC for the parameters estimation \cite{andrieu2010particle}. 	

\noindent To formulate the problem more precisely, let:
\begin{align}
&  \boldsymbol{\Theta} = \big( (\beta_{1},\ldots,\beta_{j},\ldots,\beta_{J}), (\kappa_1,\kappa_2) \big)\\
& \mathbf{Z} = \big( (z_{1,0},\ldots,z_{1,N}),\ldots,(z_{i,0},\ldots,z_{i,N}),\ldots,(z_{I,0},\ldots,z_{I,N})\big)
\end{align}	
\noindent be the arrays representing all the $J\times 2$ unknown parameters and $I\times N$ unobserved states that characterize the model's behavior. In this context, $\boldsymbol{\sigma}^2_{I}$ can be set to $\boldsymbol{1}_I$ without loss of model adequacy.\footnote{Indeed, the constraint $\boldsymbol{\sigma}^2_{I} = \boldsymbol{1}_I$ still guarantees the mapping $\pi_{ijn}:\mathbb R \to [0,1]$ to cover the needed time-to-time variability of the random walk, as Eq. \eqref{eq1_2} acts as a shrinkage operator on the support of the r.vs $\{Z_{i,0},\ldots,Z_{i,n}\}$. This has also been confirmed by several pilot simulations we ran on our model. Note that this assumption is not overly limiting, since our state-space representation is built under the \textit{smoothness assumption} on the movement behavior of the hand, according to which large abrupt changes in the small interval $[n,n+1]$ are not allowed \cite{yu2007mixture}.} The joint log-density of the complete-data given the array of parameters and the observed data is defined as follows:
\begin{align}
\log f(\mathbf{Z},\mathbf{Y}|\boldsymbol{\Theta})  & = \log f(\mathbf{Y}|\mathbf{Z},\boldsymbol{\Theta}) + \log f(\mathbf{Z}|\boldsymbol{\Theta})\label{eq3_1}\\[0.2cm]
& =\sum_{i=1}^I \sum_{j=1}^J \log f(\mathbf{z}_{i,0:N}|\boldsymbol{\Theta}) + \log f(\mathbf{y}_{ij,0:N}|\mathbf{z}_{i,0:N},\boldsymbol{\Theta})\label{eq3_2}\\[0.2cm]
& = \sum_{i=1}^I \sum_{j=1}^J \biggl( \log f(z_{i,0}|\theta_{Z_0}) + \sum_{n=1}^N \log f(z_{i,n}|z_{i,n-1},\theta_Z) + \nonumber\\
& + \sum_{n=1}^N \log f(y_{ijn}|z_{i,n},\theta_Y) \biggr)\label{eq3_3}
\end{align}
\noindent where the state and measurement equations are given as in \eqref{eq2_1}-\eqref{eq2_2} whereas the term $f(z_{ij0}|\theta_{Z_0})$ is the a-priori density function for the initial state of the process. Note that the factorization \eqref{eq3_3} is due to the Markovian properties of the model. By adopting the Bayesian perspective, we perform inference conditional on the observed sample of angles $\mathbf{Y}$, with $\boldsymbol{\Theta}$ being an unknown term. The posterior density $f(\mathbf{Z},\boldsymbol{\Theta}|\mathbf{Y})$ of hidden states and parameters is as follows:
\begin{align}
\log f(\mathbf{Z},\boldsymbol{\Theta}|\mathbf{Y})  & \propto \log f(\boldsymbol{\Theta}|\mathbf{Y}) + \log f(\mathbf{Z}|\mathbf{Y}) + \log f(\boldsymbol{\Theta})\label{eq4_1}\\
&= \sum_{i=1}^I \sum_{j=1}^J \log f(\boldsymbol{\Theta}|\mathbf{y}_{ij,0:N}) + \nonumber \\
&+ \sum_{i=1}^I \sum_{j=1}^J \log f(\mathbf{z}_{i,0:N}|\mathbf{y}_{ij,0:N}) + \log f(\boldsymbol{\Theta})\label{eq4_2}
\end{align}

\noindent where $f(\boldsymbol{\Theta})$ is a prior density ascribed on the vector of model's parameters $\boldsymbol{\Theta}$. {Note that Equation \eqref{eq4_1} comes from the standard conditional definition $f(\mathbf{Z},\boldsymbol{\Theta}|\mathbf{Y}) = f(\mathbf{Z},\boldsymbol{\Theta},\mathbf{Y}) \big/ f(\mathbf{Y})$, where the joint density $f(\mathbf{Z},\boldsymbol{\Theta},\mathbf{Y})$ is re-arranged by factorization using the Markovian properties of the model \cite{andrieu2010particle}}. Since our aim is to get samples from the posterior $f(\mathbf{Z},\boldsymbol{\Theta}|\mathbf{Y})$, we proceed by jointly updating $\boldsymbol\Theta$ and $\mathbf{Z}$ using a marginal Metropolis-Hastings. This alternates between proposing a candidate sample $\boldsymbol{\Theta}^{(t)}$ given $\boldsymbol{\Theta}^{(t-1)}$ and filtering the sequences $\mathbf{Z}^{(t)}$ conditioned on $\boldsymbol{\Theta}^{(t)}$. Finally, the candidate couple $\bigl(\boldsymbol{\Theta}^{(t)},\mathbf{Z}^{(t)}\bigr)$ is jointly evaluated by the Metropolis-Hasting ratio. 

The evaluation of both the densities $f(\mathbf{Z}|\mathbf{Y})$ and $f(\boldsymbol{\Theta}|\mathbf{Y})$ involve computing the expression in Eq. \eqref{eq4_2}. To do so, we derived the first term by means of filtering and smoothing procedures \cite{jazwinski2007stochastic} whereas the second term was evaluated by implementing a Metropolis-Hasting algorithm. {All the technical steps for the model identification are included in Appendix A-C}.

\subsection{{Model evaluation}}

The state-space model formulated can be evaluated in many ways under the Bayesian framework of analysis \cite{gelman2014bayesian,shiffrin2008survey}. For instance, adequacy of the algorithm can be assessed via standard diagnostic measures, such as traceplot of the chains, autocorrelation measures, and the Gelman-Rubin statistics whereas the recovery of the true model structure can be done by simulations from the priors ascribed to the model \cite{gelman2014bayesian}. Similarly, the adequacy of the model to reproduce the observed data can be assessed by means of simulation-based procedures (e.g., posterior predictive checks) where the fitted model is used to generate new simulated datasets that are then compared to the observed data \cite{gelman1996posterior,cook2006validation}. In our context, the robustness of the model formulation in recovering the true model structure as well as the goodness of fit to the observed data are assessed by adopting a simulation-based approach. Technical details on this procedure are available in Supplementary Materials.

\section{Application}
	
In this section, we will present an application of the model to the analysis of an already published lexical decision dataset \cite{barca2012unfolding}. The state-space modeling framework will be evaluated via three different instances of model representation with an increasing level of complexity. Note that the application we report here has only an illustrative purpose with the main goal to introduce and highlight the interpretation of the model's parameters and the flexibility of its representation with dynamic data. {All the models were estimated using 20 (chains) $\times$ 10000 (iterations), with a burning-in period of 2500 iterations. Starting values $\boldsymbol{\theta}_0$ for the MH algorithm were determined by maximizing the observed likelihood of the model in Eq. \eqref{eq2_2}. Similarly, the starting covariance matrix $\boldsymbol{\Sigma}^{(0)}$ was computed by using the Hessian of the observed likelihood at $\boldsymbol{\theta}_0$. The adaptive phase of the MH algorithm was performed at fixed interval $t+H$ (with $H=25$) to prevent the degeneracy of the adaptation. For each model, the prior densities were defined as $f(\boldsymbol \theta) = \mathcall{N}(\boldsymbol\mu=\mathbf 0,\mathbf 1\sigma^2=25)$, where the variance was sufficiently large to cover the natural range of the model parameters. The adequacy of the model to reproduce the data was evaluated with a simulation-based approach, where a series of $M=5000$ new datasets $(\mathbf{Y}^*_1,\ldots,\mathbf{Y}^*_M)$ were generated through the fitted model and compared with the observed data $\mathbf{Y}$ \cite{cook2006validation}. The goodness of fit was evaluated \textit{overall} (i.e., the adequacy of the model to reproduce the complete observed matrix $\mathbf{Y}$) and \textit{subject-based} (i.e., the adequacy of the model to reproduce for each subject $i=1,\ldots,I$ the observed matrix $\mathbf{Y}_{i}$). Comparisons were computed by means of 0-100\% normalized measures, with 0\% indicating bad fit and 100\% optimal fit. Technical details as well as extended graphical results are included in Supplementary Materials.
}

\subsection{General context and motivation}

Lexical decision is one of the most known and widely used task to study visual world recognition and reading in the cognitive psycholinguistic literature \cite{hawkins2012decision,norris2008perception,yap2008additive}. Generally, this task is very simple and versatile and provides an ideal context for applying the state-space modeling framework when lexical decision data are collected via the mouse tracking paradigm. In this application, we evaluated the extent to which the parameters of the state-space model reflect eventual differences associated with the manipulation of a stimulus type factor composed by words (with either high-frequency or low-frequency) and random strings (i.e., random sequence of letters that are phonotactically illegal in the language) in the lexical decision task. Moreover, we will take advantage of this psycholinguistic case study to show how our state-space model can deal with both categorical and (pseudo)quantitative predictive variables considered either individually or in interaction in the model. In particular, the first model instance will illustrate the application of our modeling framework when a simple categorical variable (stimulus type factor) is considered to affect the observed mouse-tracking trajectories collected using the lexical decision task. By contrast, the second model will be based on a simple regression-type model with a single quantitative independent variable (bigram frequency) used to predict the attraction toward the distractor category. Finally, the third model will integrate these two variables (stimulus type factor and bigram frequency) into a unified model including the main effects of the two variables as well as their interaction. In our context, the first two models will be considered as simple toy examples to illustrate the main features of the state-space model representation when applied to real data, whereas the third model will be discussed in more details according to a group analysis evaluation as well as an individual analysis representation.

\subsection{Model 1}

\noindent\textit{Data structure and variables}. In the original work by \cite{barca2012unfolding}, the lexical decision experiment was run in Italian and based only on one stimulus type factor with four different levels: Words of high written frequency (HF, e.g., acqua ``water"), words of low written frequency (LF, e.g., cervo ``deer"), pseudowords (PW, e.g., ``dorto"), and strings of letters that are orthographically illegal in Italian (NW, e.g., ``btfpr"). In their study, participants saw a total of 96 stimuli, one at the time, and were required to categorize each stimulus as either a word or a nonword by using the mouse-tracking paradigm. {Trajectories were recorded using the Mouse Tracker software \cite{freeman2010mousetracker} with sampling rate of approximately 70Hz \cite{barca2012unfolding}. As usual, raw trajectories were normalized  into $N=101$ time steps using linear interpolation with equal spaces between coordinate samples.} However, for our analysis we preferred to select only three of the four levels of the experimental factor (that is to say, HF,LF, and NW) for a total of 72 stimuli equally distributed within each level.\footnote{The motivation for this selection was due to some technical reasons regarding the lack of design balance in the original dataset, as the PW level showed a large number of errors when compared with the other three categories. In addition, the three-level representation of the stimulus type factor simplifies the interpretation of the results when we consider the full model with interaction. } Finally, the dependent variable $\mathbf Y$ of Model 1 consisted of the movement angles array associated with the mouse-movement trajectory recorded for each distinct stimulus in the stimulus set. \\

\begin{figure}
	\centering
	\includegraphics[scale=0.45,angle=0]{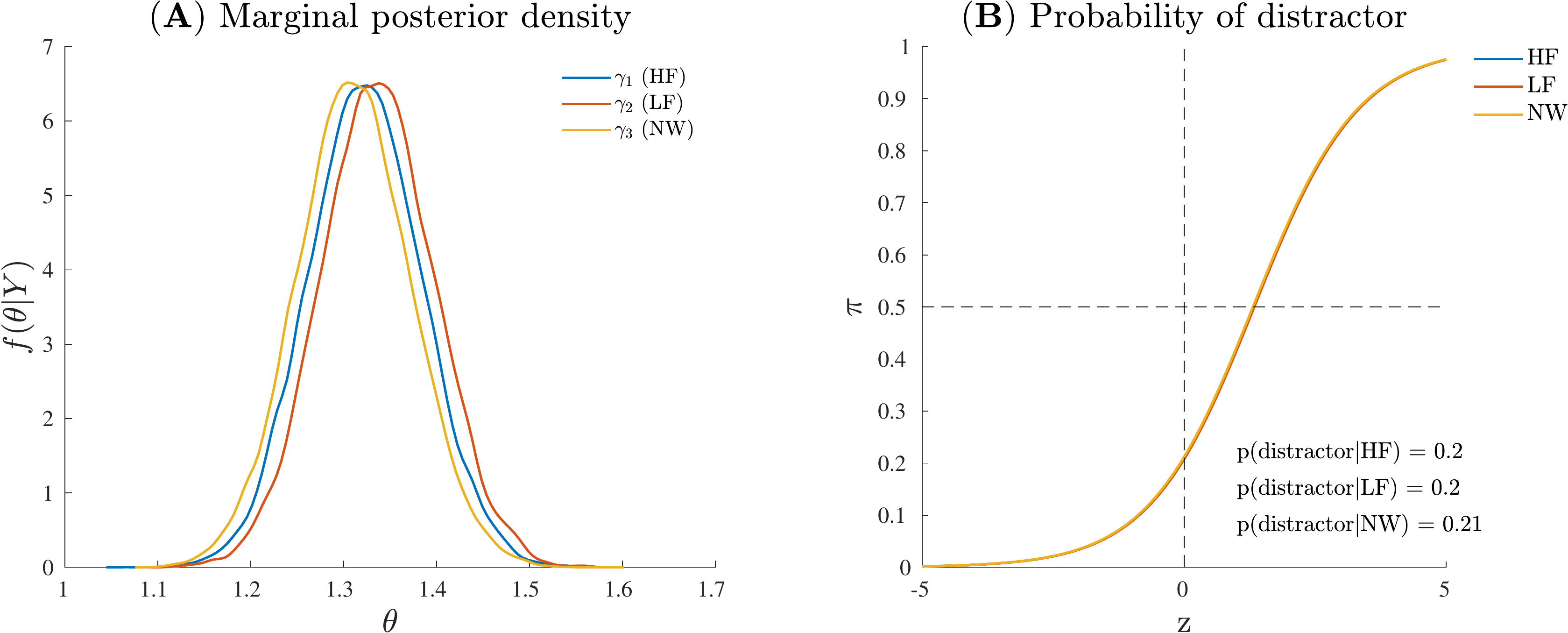}
	\caption{Application (Model 1): (A) Marginal posterior densities for the model parameters and (B) Probability to activate the distractor cue as a function of the levels (HF, LF, NW) of the categorical variable. Note that the densities in panel (A) are shown together for the sake of comparison.}
	\label{fig_mod_2}
\end{figure}

\noindent\textit{Data analysis and results}. In this first model the term $\beta_j$ in the stimuli equation boils down to the simple expression:	
\begin{eqnarray*}
	\beta_j & = & \sum_{k=1}^3 d_{jk}\gamma_k
\end{eqnarray*}
where the indices $k=1,2,3$ refer to HF, LF, and NW stimuli. The MCMC convergences of the algorithm are reported in Supplementary Materials. {The model fitted the data very satisfactorily, with overall fit of 84\% and subject-based fit of 74\% (see Table \ref{tab_fits}).} The posterior quantiles (5\%,50\%, and 95\%) are reported in Table \ref{tab_models} whereas figure \ref{fig_mod_2}-A shows the probability graph, that is to say, the probability to activate the distractor cue for each of the three levels HF, LF, NW as a function of the latent variable $Z$.

The results of this first analysis clearly show that the dynamics of the state-space model were unaffected by the different categories represented in the recoded experimental factor. This pattern finds further support in the post-hoc comparisons between the three experimental conditions (Figure \ref{fig_mod_2}-B). In sum, these findings indicate that for a dynamic model represented according to a state-space modeling framework, the three stimulus categories (HF, LF, and NW) were all processed in a very similar way, as the original trajectories were not sufficiently different among the three stimulus categories. {In substantive terms, the results of the categorical model showed how the attraction probability toward the distractor was definitively modest in all the three experimental conditions. This is evident from a direct inspection of figure \ref{fig_mod_2}-B where the probability activation function (logistic function) is shifted towards right ($Z >0$) which in turn means that the average activation of the distractor category was relatively poor in HF, LF, and NW items. In this respect, the results of our simple spatial model were partially at odds with the outcomes observed using temporal measures (response time variables) \cite{barca2012unfolding}}.

\subsection{Model 2}

\noindent\textit{Data structure and variables}. Also for the second model, the dependent variable was represented by the movement angles array $\mathbf Y$. However, unlike model 1, in model 2 the original independent categorical variable (stymulus type factor) was replaced with a quantitative psycholinguistic variable called \textit{bigram frequency}. Bigram frequency is defined as the frequency with which adjacent pairs of letters (bigrams) occur in printed texts; for its characteristics, it may be considered as a measure of orthographic typicality \cite{hauk2006time}. In this second application, only bigram frequency was used as quantitative variable, since it was the only psycholinguistic variable that could be computed for all the 72 stimuli in the stimulus set. This second model instance nicely provides a simple but effective example of application of our state-space model when a continuous variable is considered to predict the attraction toward distractor.\\

\begin{figure}
	\centering
	\includegraphics[scale=0.35,angle=0]{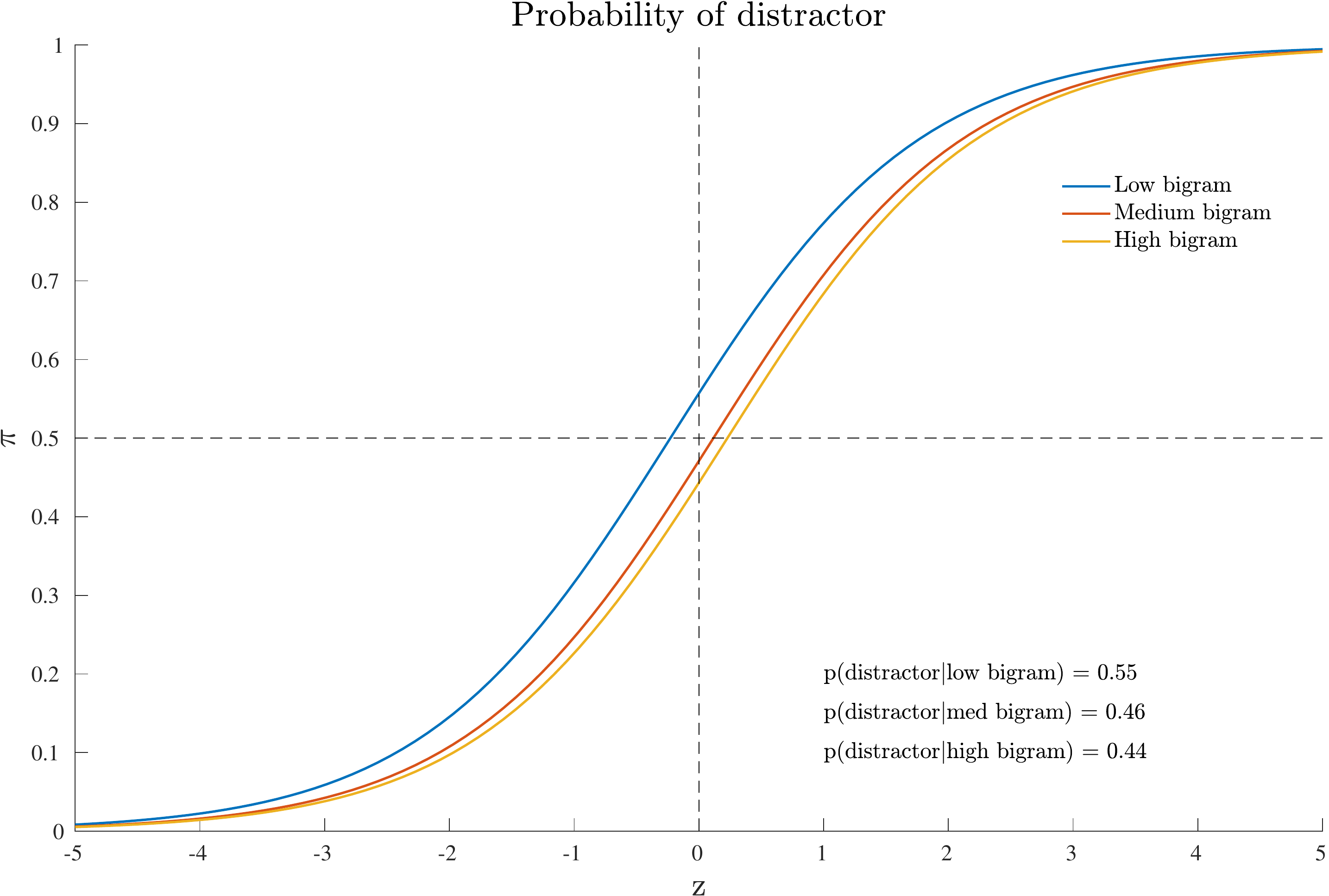}
	\caption{Application (Model 2): Probability to activate the distractor cue as a function of the continuous variable. For Note that just three representative levels (low, middle, high) are represented for the sake of graphical interpretation.}
	\label{fig_mod_4}
\end{figure}

\noindent\textit{Data analysis and results}. In model 2 the term $\beta_j$ simply reduces to:
\begin{eqnarray*}
	\beta_j & = & x_j\eta
\end{eqnarray*}
as the first and third terms in formula \eqref{eq1_4} cancel out. In this case, the variable $x_j$ denotes the value of the bigram frequency for stimulus $j$ in the stimulus set. For the model results, the posterior quantiles are reported in Table \ref{tab_models} whereas MCMC convergences of the algorithm are reported in Supplementary Materials. {Also in this case, the model fitted the data very well, with overall fit of 73\% and subject-based fit of 70\% (see Table \ref{tab_fits}).} Figure \ref{fig_mod_4} shows the probability graph for model 2. This graph represents the probability to activate the distractor hemispace at three representative levels of the variable, i.e. the lowest, the medium, and the highest values of bigram. As evident from the graph, bigram frequency affects the probability to activate the target, with a higher probability for stimuli with low bigram frequency.

{In substantive terms, the results of the quantitative model supported the evidence that the attraction probability toward the distractor was slightly affected by the specif value of the quantitative predictor (bigram frequency). In particular, low-level bigram frequencies were characterized by an average larger activation probability ($0.55$) for the distractor, whereas medium or large frequencies were associated with a logistic function slightly shifted toward positive values of the latent space $Z>0$, thus reflecting a lower chance for the distractor category (average activation probability of $0.45$). Moreover, by an inspection of the contingency table for the joint representation of bigram frequency (as a transformed categorical variable) and stimulus type, we noted that low bigram frequency values were mainly characterized by string letters (NW: $94\%$) whereas high bigram frequency values were predominantly associated with high frequency words (HF: $55\%$) or low frequency words (LF: $44\%$)}.\\

\subsection{Model 3}
\noindent\textit{Data structure and variables}. The final and more complex model included both the three-level categorical predictor (stimulus type factor: HF, LF, STR) and the continuous predictor (bigram frequency) as well as the interaction term between these two variables. The dependent variable was the movement angles array $\mathbf Y$.\\

\begin{figure}
	\centering
	\includegraphics[scale=0.42,angle=0]{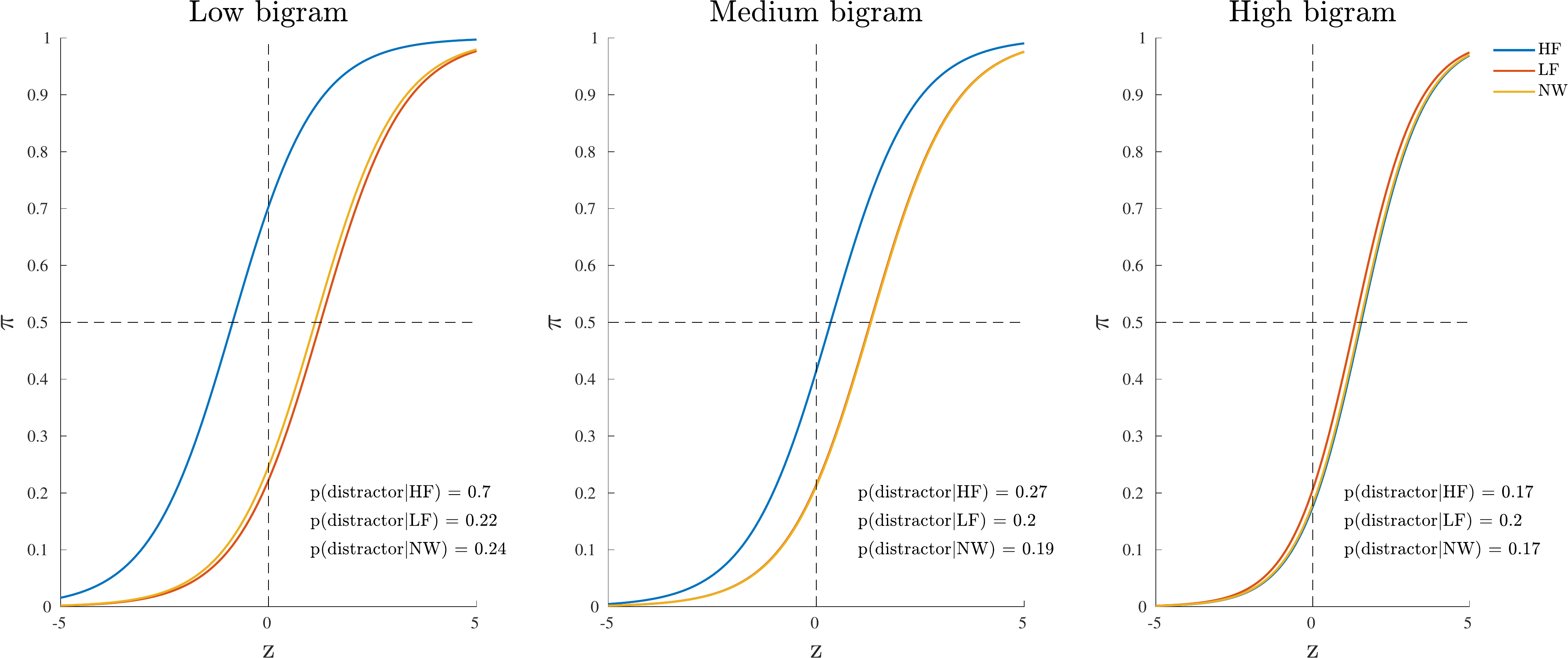}
	\caption{Application (Model 3): Probability to activate the distractor cue as a function of the categorical variable (within panels) and three representative levels of the continuous variable (between panels). }
	\label{fig_mod_7}
\end{figure}

\noindent\textit{Data analysis and results}. The stimuli equation which characterizes the third model is defined as follows:	
\begin{eqnarray*}
	\beta_j & = & \sum_{k=1}^3 d_{jk}\gamma_k + x_j\bigg(\eta + \sum_{k=1}^3 d_{jk}\delta_k\bigg)
\end{eqnarray*}
The MCMC diagnostics together with the  estimated marginal posterior densities for the model's parameters are reported in Supplementary Materials. {The model fit was good, with an overall fit of 75\% whereas the subject-based fit was equal to 71\% (see Table \ref{tab_fits}).} The posterior quantiles are reported in Table \ref{tab_models}. Figure \ref{fig_mod_7} shows the probability graph for model 3. This graph represents the probability to activate the distractor hemispace for each of the three levels HF, LF, NW of the categorical factor as a function of the latent variable $Z$ and three distinct levels (high, medium, and low) for bigram frequency. The inspection of Figure \ref{fig_mod_7} shows a clear interaction between stimulus type factor and bigram frequency indicating that the impact of stimulus category, in particular word frequency, increases with the decrease of stimulus bigram frequency. In other words, at high level of bigram frequency, the probability to activate the distractor is similarly low in all conditions (.17 $\leq$ p-distractor $\leq$ .2). By contrast, when bigram frequency decreases - that is stimuli become orthographically atypical - the probability of distractor activation increases, but only for the more lexically-familiar stimuli, i.e., words of high frequency (e.g., p-distractor raises from 0.17 to 0.70, in low and high bigram frequency condition respectively).

\begin{figure}
	\centering
	\includegraphics[scale=0.45,angle=0]{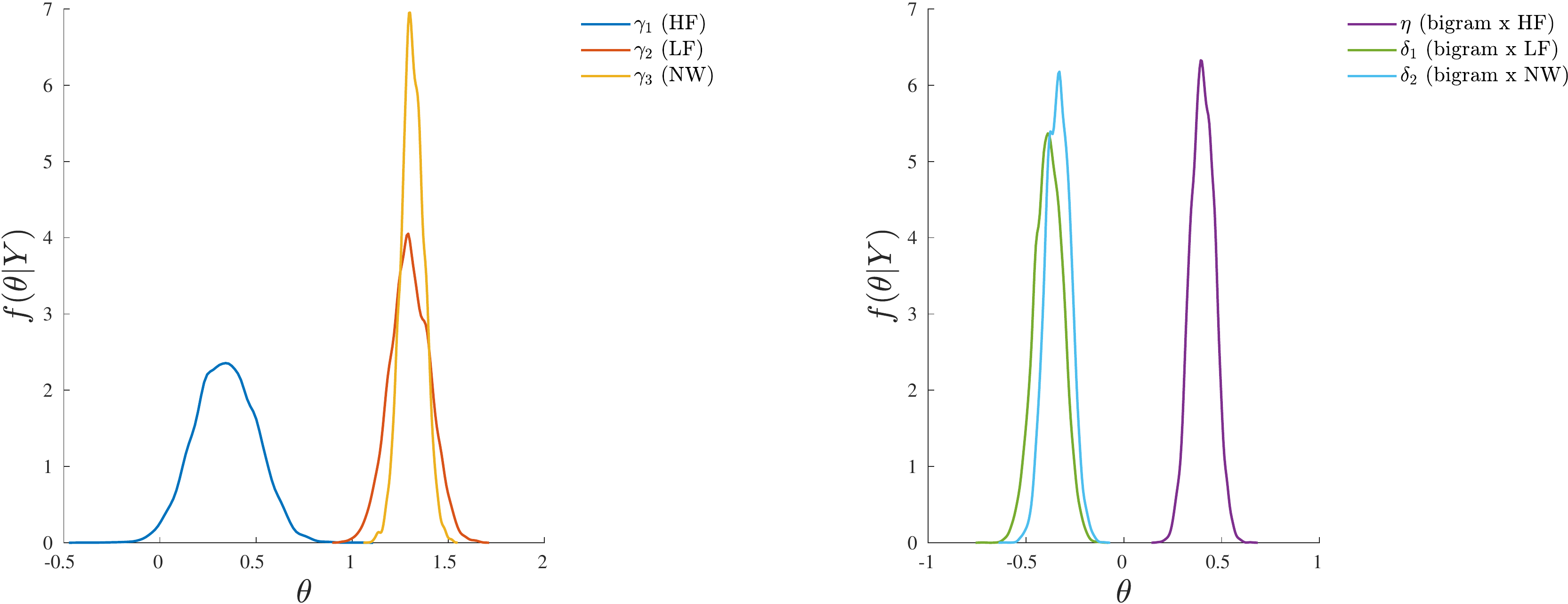}
	\caption{Application (Model 3): Marginal posterior densities for the model parameters. Left panel: parameters associated to the categorical variable. Right panel: parameters associated to the continuous variable and its interaction with the categorical variable. Note that the densities are shown together for the sake of comparison.}
	\label{fig_mod_6}
\end{figure}

Finally, it is also worth mentioning the emergence of the main effect of stimulus category which was instead missing in model 1. By a quick inspection of Figure \ref{fig_mod_6}, one may clearly observe that HF words differ from both LF words and letter strings (NW), whereas LF words and letter strings do not differ with respect to the probability of activation of the distractor hemispace. Interestingly, the addition of the covariate bigram frequency in the model allowed the main effect of stimulus category to show up. Indeed, while at the medium and high levels of bigram frequency the results are in line with those observed at a sample level in the original study (see Figures 1,2,5 in \cite{barca2012unfolding}) and in a recent re-analysis (see Table 2 in \cite{calcagni2017analyzing}), in the case of low bigram the probability to activate the distractor increases with respect to high frequency words (HF). This might be somewhat related to a moderate difficulty in the orthographic processing of low frequency bigram words \cite{rastle2008morphological} even in the case of stimuli with richer lexical representation.

\begin{table}[!h]
	\centering
	\begin{tabular}{lcc}
		\toprule[\heavyrulewidth]\toprule[\heavyrulewidth]
		& overall & by-subject \\\midrule 
		Model 1 & 84\% & 78\% \\
		Model 2 & 73\% & 70\% \\
		Model 3 & 75\% & 71\% \\
		\bottomrule[\heavyrulewidth]\bottomrule[\heavyrulewidth]
	\end{tabular}
	\caption{Application: Adequacy of the model to reproduce the observed matrices $\mathbf Y$ (\textit{overall fit}) and $\mathbf Y_i$ (\textit{by-subject fit}). All the measures are normalized in the range 0\% (bad fit) - 100\% (otimal fit). See Supplementary Materials for technical details.}  
	\label{tab_fits}
\end{table}

\begin{table}[!h]
	\centering
	\begin{tabular}{lllcccccc} 
		\toprule[\heavyrulewidth]\toprule[\heavyrulewidth]
		& &  & $\hat\gamma_1$ & $\hat\gamma_2$ & $\hat\gamma_3$ & $\hat\eta$ & $\hat\delta_1$ & $\hat\delta_2$ \\\midrule 
		& & & & & & & \\ 
		\multirow{5}{*}{Model 1} 
		& & $q_{0.05}$ & 1.224 & 1.234 & 1.211 & & &\\[0.1cm]
		& & $\mu$ & 1.323 & 1.337 & 1.310 & & & \\[0.1cm]
		& & $q_{0.975}$ & 1.443 & 1.457 & 1.432 & & &\\[0.1cm]
		& & $\hat R$ & 1.003 & 1.002 & 1.003 & & &\\[0.5cm]
		\multirow{5}{*}{Model 2} 
		& & $q_{0.05}$ &  &  & & 0.063 & & \\[0.1cm]
		& & $\mu$ &  &  & & 0.078 & & \\[0.1cm]
		& & $q_{0.975}$ &  & & & 0.091 & & \\[0.1cm]
		& & $\hat R$ &  & & & 1.001 & &\\[0.5cm]
		\multirow{5}{*}{Model 3} 
		& & $q_{0.05}$ & 0.083 & 1.130 & 1.217 & 0.305 & -0.505 & -0.437\\[0.1cm]
		& & $\mu$ & 0.341 & 1.300 & 1.314 & 0.402 & -0.385 & -0.336\\[0.1cm]
		& & $q_{0.975}$ & 0.605 & 1.468 & 1.411 & 0.500 & -0.269 & -0.235\\[0.1cm]
		& & $\hat R$ & 1.008 & 1.013 & 1.012 & 1.011 & 1.013 & 1.010\\[0.5cm]
		\multirow{2}{*}{All the models}
		& & $\hat\kappa_1 = 22.31$ & & & & & & \\
		& & $\hat\kappa_2 = 44.96$ & & & & & & \\[0.25cm]		
		\bottomrule[\heavyrulewidth]\bottomrule[\heavyrulewidth]
	\end{tabular}
	\caption{Application: Posterior means ($\mu$), 95\% posterior intervals ([$q_{0.05},q_{0.975}$]), and Gelman-Rubin $\hat R$ index for the estimated parameters of Models 1-3. }  
	\label{tab_models}
\end{table}

\subsection{Profiles analysis}

\begin{figure}
	\centering
	\includegraphics[scale=0.45,angle=0]{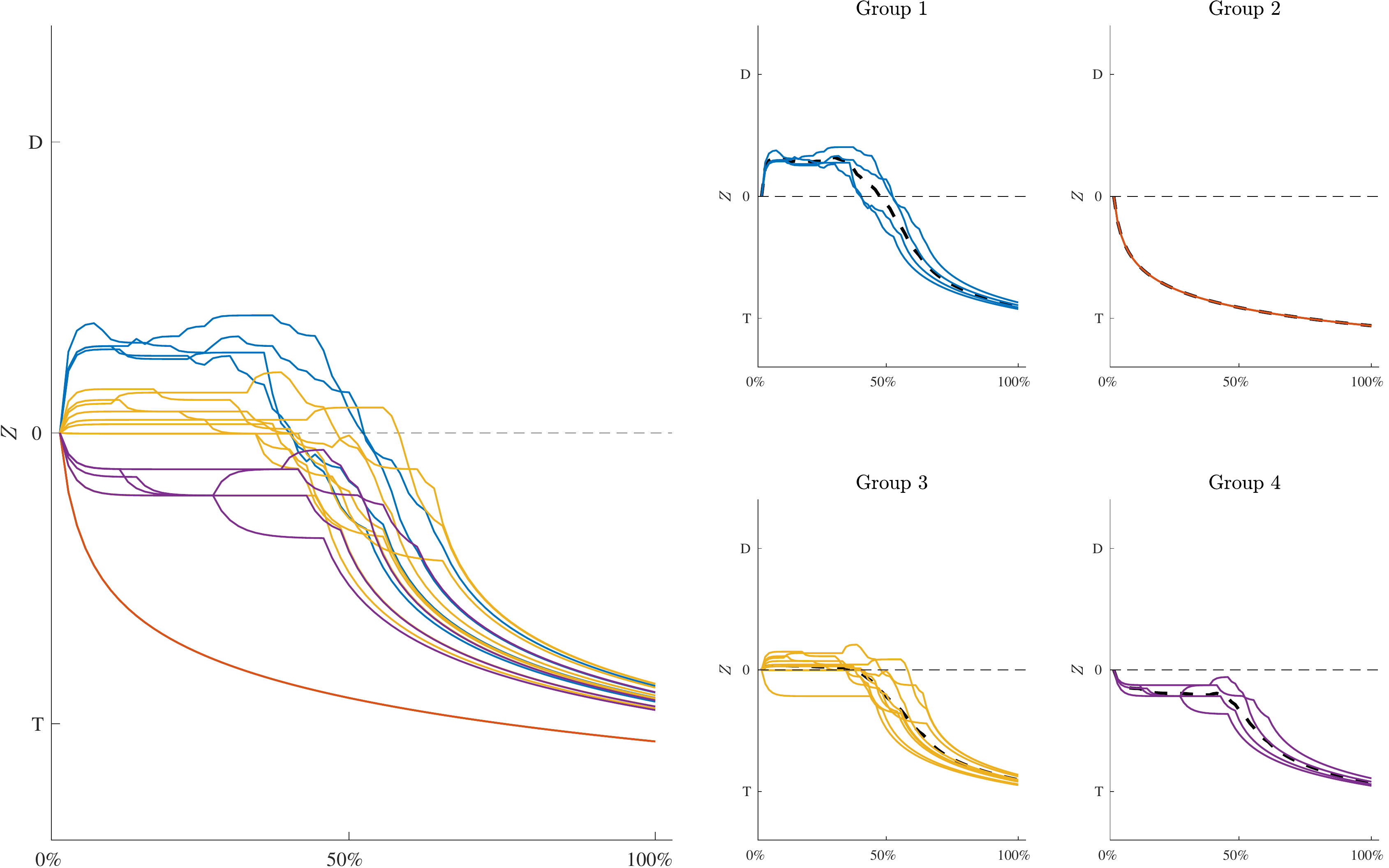}
	\caption{Application: Estimated movement dynamics $\mathbf z_{i,0:N}$ of each participant (biggest panel, left) and clusters of profiles in terms of their functional similarity (smallest panels, right). Note that averaged profiles are represented as dashed lines whereas D and T in all the panels indicate distractor and target, respectively. Groups' composition: participants 6,7,8,15 (group 1), 1,4,19,21 (group 2), 2,3,5,12,13,16,17,20,22 (group 3), 10,11,14,18 (group 4). }
	\label{fig2p_1}
\end{figure}

\begin{figure}
	\hspace*{-0.8cm}
	\includegraphics[scale=0.5,angle=0]{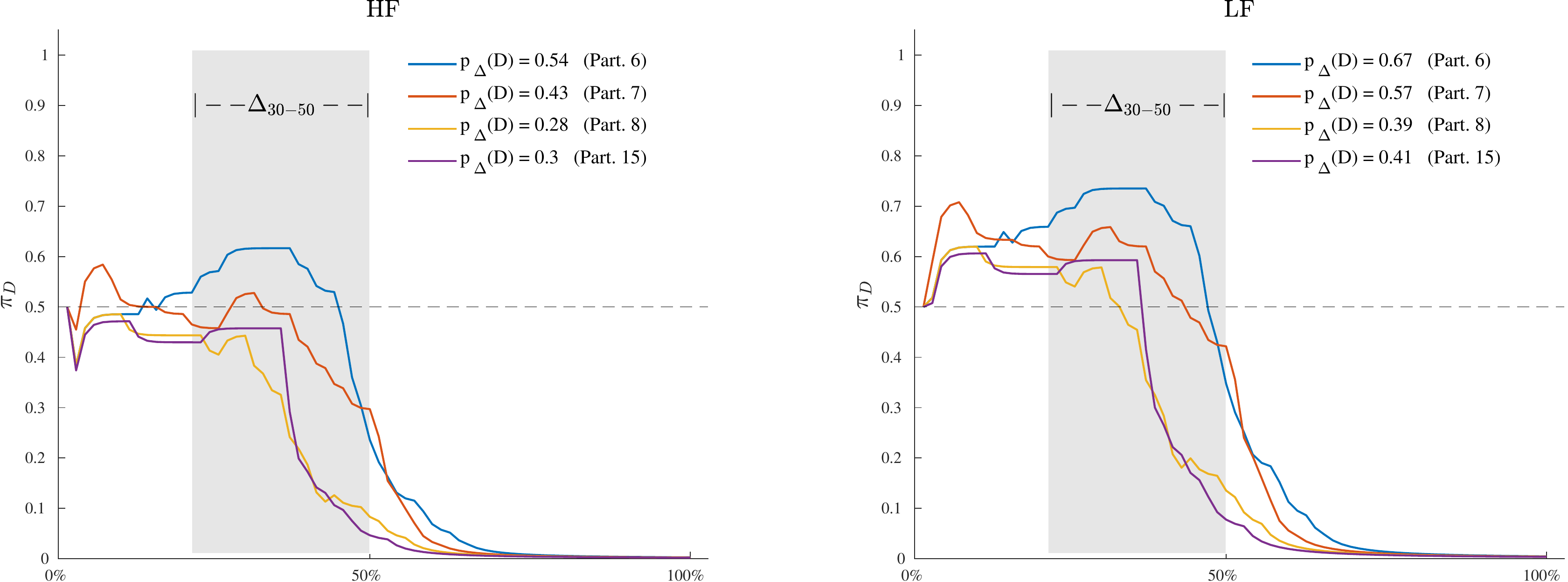}
	\caption{Application: Estimated attraction probabilities $\boldsymbol \pi_{i,0:N}$ of participants in Group 1 for the High Frequency (left panel) and Low Frequency (right panel) lexical conditions. Note that the probability curves are computed with respect to the distractor (D), the gray area in both panels indicates a selected window of processing ($\Delta=30\%-50\%$), whereas the terms $p_\Delta(D)$ are computed using a normalized discrete approximation of the integral of the probability curves in the selected process window $\Delta$.}
	\label{fig2p_2}
\end{figure}

\begin{figure}
	\hspace*{-2.5cm}
	\includegraphics[scale=0.425,angle=0]{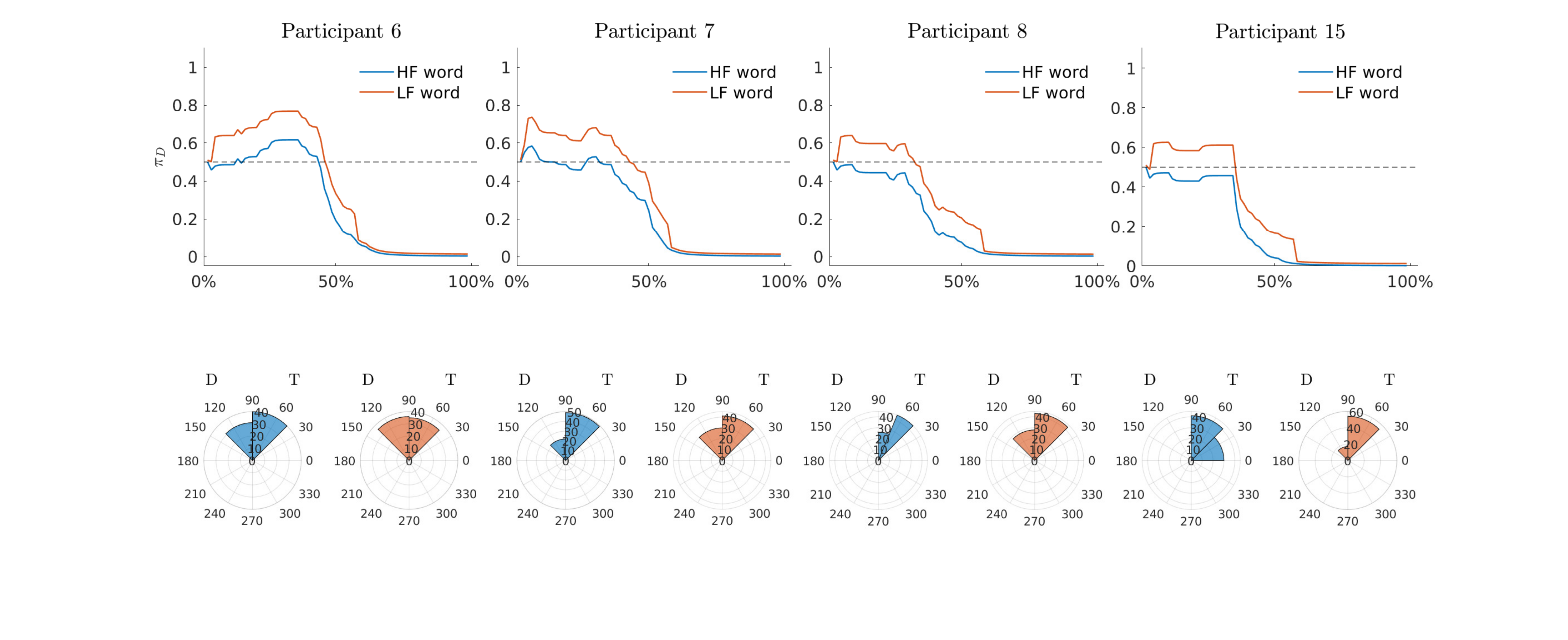}
	\caption{Application: Estimated attraction probabilities $\boldsymbol \pi_{i,0:N}$ of participants in Group 1 and rose diagrams of observed radians for two stimuli (HF: \textit{epoca}, epoch. LF: \textit{zampa}, paw). Note that D and T in all the panels indicate distractor and target, respectively.}
	\label{fig2p_3}
\end{figure}

To further investigate the dynamic characteristics involved in the lexical decision task, we extend here the results of the third model to include also a profiles analysis. Figure \ref{fig2p_1} shows the estimated latent movement states $\mathbf Z_{I\times N}$ for all the participants involved in the study. The profiles appear regular, as they evolve smoothly toward the target cue (T). We grouped the dynamics into four well-separated clusters (Figure \ref{fig2p_1}, smallest panels on the right) according to their functional similarities \cite{ramsay2006functional}. Particularly, the first group is characterized by a higher exploration of the distractor's hemispace, especially in the first 30\% of the process. The same applies to the third and fourth groups, although they show a gradual activation of the distractor. Finally, the second group clearly represents those profiles with no uncertainty in the categorization process, as they show no activation of the distractor's hemispace at all. Although well-separated among them, these clusters still show some level of inner heterogeneity (for example, see group 1 and 4). To study this latter issue in terms of experimental manipulations, we focused on group 1 and considered the low vs. high frequency conditions (HF vs. LF). We also selected the middle phase of the process ($\Delta=30\%-50\%$), where it is expected to observe larger cognitive competitions in the categorization \cite{barca2012unfolding}. Figure \ref{fig2p_2} shows the participants' profiles in terms of attraction probability $\boldsymbol \pi_{4\times N}$ for the two lexical conditions. As expected, the profiles differ between these conditions, with LF eliciting higher attraction probability. This is in line with the fact that low frequency words have a weaker lexical representation than high frequency stimuli and consequently they are more difficult to process \cite{barca2012unfolding}. Interestingly, the individual profiles also differ in the way they activate the distractor. For instance, the participant 6 had higher probability in both LF ($p_\Delta(D) = 0.67$) and HF ($p_\Delta(D) = 0.54$) conditions whereas the participant 7 had a more pronounced activation just in the LF condition ($p_\Delta(D) = 0.57$) than HF ($p_\Delta(D) = 0.43$). Similarly, participants 6 and 7 seemed to prolong the competing dynamics up to the 50\% of the process, by contrast participants 8 and 15 seemed to resolve the lexical competition earlier as showed by the abrupt decreasing of their curves. We complete our analysis by evaluating how individual profiles are linked to empirical measurements. Figure \ref{fig2p_3} represents this scenario for two stimuli belonging to HF and LF conditions. As we can notice, the curves present the same dynamics (due to the latent states $\mathbf z_{i,0:N}$) although they clearly differ in terms of attraction exerted by the stimulus (due to the $\beta$ component of the model). In this case, the LF stimulus produced larger conflict than HF in the lexical categorization. This is evident when we turn back to the observed data: as expected, the rose diagrams of LF showed larger directions in the distractor's hemispace.  

\section{Discussion}

We have described a new approach to model and analyse dynamic data coming from mouse-tracking experiments. Our proposal took the advantages of a state-space representation, in which the observed data $\mathbf Y$ were thought as being function of two independent sub-models, one representing the movement process and its properties ($\mathbf Z$) and the second modeling the two-choice experimental task ($\boldsymbol\beta$) according to which the data were collected. These sub-models were integrated by means of an inverse-logit function ($\pi$) that expressed how the experimental manipulations acted on the movement processes in selecting the final correct response against the competing one. This formulation was flexible enough to take into account the complexity of some dynamic behaviors showed by the reaching trajectories. Moreover, it allowed for separately accounting for the motor heterogeneity and experimental variability in $\mathbf Y$. Indeed, when $\boldsymbol\beta = \mathbf 1\beta_0$ our state-space representation simply reduced to a model where the experimental manipulations had no relevant effect in reproducing the observed data. This instance has been illustrated in Section 4.3 (Model 1). In this case, as $\mathbf{Z}=\mathbf 0$ was not allowed in our model formulation, all the variability of $\mathbf Y$ can be ascribed to $\mathbf Z$. This is relevant in view of the fact that movement variability may reflect only individual motor executions in absence of any experimental manipulations \cite{yu2007mixture}. The movement component $\mathbf Z$ was modeled to be Markovian with gaussian transition density.

Although more complex models can be used to represent movement dynamics, simple random walks still allows a great deal of flexibility in modeling reaching trajectories under weak assumptions on the movement behavior \cite{yu2007mixture,paninski2010new}. In particular, in the case of mouse-tracking tasks, they allow representations of the following three properties: (i) Each movement is goal-oriented as individuals have to finalize the action by clicking on one of the two categories shown on the screen. (ii) Mouse-tracking trajectories generally start at rest, proceed out in the movement space, and end at rest. (iii) Hand trajectories tend to be smooth during the reaching process, i.e. small changes in the interval $[n,n+1]$ are more likely than large and abrupt changes \cite{brockwell2004recursive,spivey2010curved}. The stimuli component $\boldsymbol{\beta}$ was defined to be a linear combination of information typically involved in a univariate design, namely a categorical variable $D$ containing the levels of the experimental factor and a continuous covariate $X$. This gave researchers the opportunity to additionally analyse which component of the experimental design is relevant in producing the effect of $\boldsymbol{\beta}$ on $\mathbf Y$. The data-generation process was defined according to a mixture of two von-Mises distributions representing the categories of a two-choice categorization task. Among others, we chose this distribution because it provides a flexible representation for angular ordered data, especially because it simplifies mathematical computations involved in the model's derivation \cite{mcclintock2012general,mulder2017bayesian}. 

{There are other existing methods that offer alternative ways to model mouse-tracking data. For instance, \cite{van2009trajectories} proposed the use of the \textit{movement superposition model} \cite{henis1995mechanisms} to model and analyse the typical two-choice lexical decision task. In particular, they modeled mouse-tracking trajectories by representing the complete hand movement as a summation of sub-movements, which were obtained by the solution of the minimum-jerk equation for the standard reaching trajectory (i.e., a movement characterized by a bell-shaped speed profile that minimizes the sum of the squared rates of jerks over the movement duration). Similarly, \cite{friedman2013linking} discussed how an intermittent model of arms movement can be used for reaching trajectories in random-dot experiments. They used both Wiener's diffusion process and Flash and Hogan's movement equation to predict reaction times (RTs) and movement data. Their goal was to assess the link between movement trajectories and underlying cognitive processing. } {Our model differs in some respects from these works. With regards to \cite{van2009trajectories}, for instance, we used a stochastic state-space approach to model the movement trajectories instead of deterministic equations. Instead, with respect to \cite{friedman2013linking}, we tailor-made our model to a typical two-choice categorization task, making use of few assumptions on the nature of the movement process (as those implied by the Gaussian AR(1) process). By and large, our goal was not to provide a mathematical representation of the cognitive components underpinning mouse-trajectories since the model does not describe the cognitive processing \textit{per se}. By contrast, we simply provided a \textit{statistical model} for the analysis of mouse-tracking data, which can offer a good compromise between data modeling and data analysis.}

\subsection{Model's advantages and limitations}
Our non-linear state-space model has several advantages. For instance, when comparing with the standard approaches, our proposal provides a unified analytic framework to simultaneously model and analyse trajectories data. By modeling movement heterogeneity and task variability together, we can evaluate how experimental variables directly act on the observed series of trajectories, with no need to use any kind of summary measures. An additional advantage of our model concerns the study of individual differences in terms of latent dynamics. While this is impractical in standard two-step approaches, in our proposal researchers can assess individual variations by studying the movement profiles $\mathbf{\tilde{Z}}$ once they are estimated. For instance, they can be analysed in terms of similarity/dissimilarity with regards to external individual covariates (e.g., vocabulary knowledge and bilingualism in psycholinguistic experiments; IQ, risk-taking propensity, or more generally clinical variables in decision-making tasks). Still, individual dynamics can be compared each other qualitatively in terms of chance to activate the distractor or target cues. As the dynamics are normalized on a common cumulative scale, researchers can assess whether the chance to activate the distractor cue at a certain percentage of the process and for an experimental manipulation, is particularly higher in a sub-group of participants (this case, for example, has been described in Section 4.6). 

As for any modeling approach, also the current proposal can potentially suffer from some limitations. A first limitation concerns the only-intercept model $\pi(\mathbf Z,\boldsymbol\beta)$ used to integrate individual dynamics and experimental information. Although this was enough to represent whether or not certain stimuli can increase the probability to select the distractor cue, we may want to known whether some experimental manipulations can modify the individual dynamics as well. However, this would particularly pronounce the computational costs required for the model identification (especially with regards to filtering), as we need to appropriately generalize Eq. \eqref{eq1_2} to include more parameters. Lastly, in the current study we used univariate non-linear state-space models to represent individual dynamics for the sake of parsimony. However, more complicated situations may require models including further movement characteristics like step-length, velocity, acceleration, and jerk \cite{kulkarni2008state}, which may be modeled as statistical constraints of the model \cite{ciavolino2014generalized,calcagni2017analyzing}.

\subsection{Further extensions}
Our non-linear state-space model can be improved in many aspects. For instance, the stimuli equation \eqref{eq1_4} can be generalized to cope with more complex experimental designs, like those involving multiple factors and covariates together with high-order interaction terms. Likewise, the current model restrictions can be relaxed to allow changes in slopes of $\pi(\mathbf Z,\boldsymbol\beta)$ as a function of the experimental stimuli. Further, the development of a hierarchical representation of the model, with a random-effect component in the state equation \eqref{eq1_2}, would offer a way to model the inter-individual variations as resulting from an underlying common population. Still, the development of a multivariate state-space model to include other movement components will surely be considered a future extension of the present work. {Further studies may lead to generalize the AR(1) process used for the movement dynamics to include former knowledge on the deterministic constraints of the hand movement as those used, for instance, by \cite{van2009trajectories} and \cite{friedman2013linking}.} {Moreover, further studies may also lead to generalize the AR(1) process used for the movement dynamics to include former knowledge on the deterministic constraints of the hand movement as those used, for instance, by \cite{van2009trajectories} and \cite{friedman2013linking}. Finally, an open issue which deserves greater consideration in future investigations is the need for a formal comparative framework with which we may eventually contrast and compare spatial modeling perspectives (like the one presented in this contribution) and currently used methods for analyzing mouse tracking data based on descriptive statistics \cite{freeman2017}.}

\section{Conclusions}
In this paper we presented a novel and comprehensive analytic framework for modeling and analyse mouse-tracking trajectories. In particular, a non-linear state-space approach was used to model the observed trajectories as a function of both individual movement dynamics and experimental variables. Model identification was performed under the umbrella of Bayesian methods, in which a Metropolis-Hastings algorithm was coupled with a recursive gaussian approximation filter to get posterior distributions of model parameters. For the sake of illustration, we applied our new approach to a real mouse-tracking dataset concerning a two-choice lexical categorization task. The results indicated how our proposal can provide valuable insights to assess the dynamics involved in the decision task and identify how the experimental variables significantly contributed to the observed movement heterogeneity. Moreover, the analysis of individual profiles allowed for comprehensive and reliable identification of individual and group-based differences in the dynamics of decision making. 

In conclusion, we think that this work yielded interesting findings in the development of computational models able to capture the unfolding high-level cognitive processes as reflected by motor executions which are typically involved in mouse-tracking tasks. To our knowledge, this is the first time that mouse-tracking data are fully modeled and analysed within a process-oriented approach. We believe our contribution will offer a novel strategy that may help cognitive researchers to understand the roles of cognition and action in mouse-tracking based experiments.

\vspace{3.5cm}
\noindent\textbf{Acknowledgments}. The authors thank Dr. Simone Sulpizio for many helpful discussions on earlier versions of the manuscript and Dr. Laura Barca for providing the dataset used in the case study. This work was supported by CARITRO, Fondazione Cassa di Risparmio di Trento e Rovereto (ref. grant 2016.0189).

\clearpage
\renewcommand{\theequation}{A.\arabic{equation}}
\setcounter{equation}{0}  
\renewcommand{\thetable}{A.\arabic{table}}
\setcounter{table}{0}  
\small
\linespread{0.35}

\section*{Appendix A: Filtering and smoothing}  

\noindent The term $f(\mathbf{Z}|\mathbf{Y})$ in Eq. \eqref{eq4_2} is recursively computed given all the measurements up to the $n$-th step. Let:
\begin{align}\label{eq5}
	\log f(z_{i,n}|\mathbf{y}_{ij,0:n},\theta)	& \propto \log f(y_{ijn}|z_{i,n},\theta) + \\\nonumber
	& + \log \int_{\mathbb R}f(z_{i,n}|z_{i,n-1},\theta)f(z_{i,n-1}|\mathbf{y}_{ij,0:n-1},\theta) dz_{i,n-1}
\end{align}
\noindent be the filtering density at step $n$. The first term of the right-side of the equation is the observation equation whereas the second term represents the Chapman-Kolmogorov equation. Given the initialization at $n=0$, the filter proceeds by solving the integral in the Chapman-Kolmogorov equation (prediction step) and compute the log posterior filtering density (update step). In our model, we solve Eq. \eqref{eq5} by means of a Gaussian approximation filter \cite{smith2003estimating}, which computes a gaussian approximation to the posterior density $f(z_{i,n}|\mathbf{y}_{ij,0:n},\theta)$ and determines its posterior mode $z_{i,n|n}$ and variance $\lambda^2_{i,n|n}$ recursively. \\
\noindent More technically, let: 
\begin{align}\label{eq5A}
	\log f(z_{i,n}|\mathbf{y}_{ij,0:n},\theta)	& \propto \log f(y_{ijn}|z_{i,n},\theta) + \\\nonumber
	& + \log \int_{\mathbb R}f(z_{i,n}|z_{i,n-1},\theta)f(z_{i,n-1}|\mathbf{y}_{ij,0:n-1},\theta) dz_{i,n-1}
\end{align}
be the filtering density at step $n$. Consider the following definitions:
\begin{align}
	& y_{ijn}|z_{i,n} ~\sim~ {mixVM}(\mu_1,\mu_2,\kappa_1,\kappa_2,\pi_{ijn})\label{eq6_1}\\[0cm]
	& z_{i,n}|z_{i,n-1} ~\sim~ \mathcall{N}(z_{i,n-1},\sigma^2_i)\label{eq6_2}\\[0cm]
	& z_{i,n-1}|\mathbf{y}_{ij,0:n-1} \sim~ \mathcall{N}(z_{i,n-1|n-1},\lambda^2_{i,n-1|n-1})\label{eq6_3}
\end{align}
\noindent where $z_{i,n-1|n-1}$ and $\lambda^2_{i,n-1|n-1}$ represent the mode and the variance of the gaussian approximation in the prediction step. Under definitions \eqref{eq6_2}-\eqref{eq6_3}, integrating out for $z_{i,n-1}$ in Eq.\eqref{eq5A} yields to:
\begin{equation}\label{eq6_4}
	z_{i,n}|\mathbf{y}_{ij,0:n-1},\theta \sim \mathcall{N}(z_{i,n-1|n-1},\lambda^2_{i,n-1|n-1}+\sigma^2_i)	
\end{equation}
\noindent For the sake of computational simplicity, we rewrite the measurement density in Eq. \eqref{eq2_2} as follows:
\begin{align}\label{eq2_2b}
	f(y_{ijn}|\pi_{ijn},\theta) &= \biggl[\frac{\pi_{ijn}}{2\pi I_0(\kappa_1)}\exp\biggl(\cos(y_{ijn}-\mu_1)^{\kappa_1}\biggr)\biggr]^{u_{ijn}} \cdot \nonumber\\ 
	& \cdot \biggl[\frac{1-\pi_{ijn}}{2\pi I_0(\kappa_2)}\exp\biggl(\cos(y_{ijn}-\mu_2)^{\kappa_2}\biggr)\biggr]^{1-u_{ijn}}
\end{align} 
where $u_{ijn}$ is a (deterministic) indicator variable taking the value 1 when $y_{ijn}$ is in the area of the screen associated to \textit{C1}, and 0 when $y_{ijn}$ is in the area associated to \textit{C2} (see Fig. \ref{fig1}) \cite{banerjee2005clustering}. Next, using these results together with Eqs. \eqref{eq6_1}-\eqref{eq6_3} we obtain:
\begin{equation}\label{eq7}
	\begin{split}
		\log f(z_{i,n}|\mathbf{y}_{ij,0:n},\theta)	&\propto \log f(y_{ijn}|z_{i,n},\theta) + \log f(z_{i,n}|\mathbf{y}_{ij,0:n-1},\theta)\\
		&= u_{ijn}\bigl(\kappa_1\cos(y_{ijn}-\mu_1) + \log \pi_{ijn}-\log I_0(\kappa_1)\bigr) + \\ 
		&+ (1-u_{ijn})\bigl(\kappa_2\cos(y_{ijn}-\mu_2) + \log (1-\pi_{ijn})-\log I_0(\kappa_2)\bigr) + \\
		&+ \frac{1}{2}\Biggl(\log (\lambda^2_{i,n-1|n-1}+\sigma^2_i) - \frac{(z_{i,n}-z_{i,n|n-1})^2}{\lambda^2_{i,n-1|n-1}+\sigma^2_i}\Biggr)
	\end{split}
\end{equation}
Differentiate $\mathcall{F} \triangleq f(z_{i,n}|\mathbf{y}_{ij,0:n},\theta)$ around $z_{i,n|n}$ gives:
\begin{align}
	\frac{\partial \log \mathcall{F}}{\partial z_{i,n|n}} =& ~\sum_{j=1}^J \Biggl( \frac{2z_{i,n|n-1}-2z_{i,n|n}}{2\sigma^2_i +\lambda^2_{i,n-1|n-1}} + \frac{\exp\big(\beta_j-z_{i,n|n}\big)u_{ijn}}{\exp\big(\beta_j-z_{i,n|n}\big)+1} -\nonumber\\
	&- \frac{\exp\big(\beta_j-z_{i,n|n}\big)(u_{ijn}-1)}{\exp\big(2z_{i,n|n}-2\beta_j\big)-1}  \Biggl)\label{eq8_1}
\end{align}
\begin{align}
	&\frac{\partial^2 \log \mathcall{F}}{\partial z_{i,n|n}^2} = ~\sum_{j=1}^J \Biggl(-\frac{1}{4\cosh\bigl(\frac{1}{2}(\beta_j - z_{i,n|n}\bigr)^2}-\frac{1}{(\sigma^2_i +\lambda^2_{i,n-1|n-1})}\Biggr)\label{eq8_2}
\end{align}
where $\cosh(x) \triangleq \bigl(1+\exp(-2x)\bigr)/\bigl(2\exp(-x)\bigr)$ is the hyperbolic cosine function. 	\\
Finally, the posterior moment $z_{i,n|n}$ is obtained by solving Eq. \eqref{eq8_1} whereas $\lambda^2_{i,n|n}$ is computed by the negative inverse of Eq. \eqref{eq8_2} \cite{tanner1991tools}. As Eq. \eqref{eq8_1} is non-linear, it can be solved numerically (e.g., using the Broyden's method). The complete filtering algorithm is summarized in Table \ref{tab1}.

\begin{table}[!h]
	\centering
	\begin{tabular}{llr}\\[0.05cm]\hline
		\small{\textit{Algorithm 1}} &{Gaussian Approximation filter algorithm}&\\[0.05cm]\hline
		$\mathbf{n=0:}$ & $z_{i,0} = 0$ & \textsc{initialization}\\[0.2cm]
		& $\lambda^2_{i,0} = 1$ & \\[0.3cm]
		$\mathbf{n>0:}$ & $z_{i,n|n-1} ~=~ z_{i,n-1|n-1}$ & \textsc{prediction}\\
		& $\lambda^2_{i,n|n-1} ~=~ \lambda^2_{i,n-1|n-1} + \sigma^2_i$ & \\[0.3cm]
		& $z_{i,n|n} ~=~ \mbox{solve}\big({\partial \log f(z_{i,n}|\mathbf{y}_{ij,0:n},\theta)}\big/{\partial z_{i,n|n}}\big)$ & \textsc{update}\\[0.3cm]
		& $\lambda^2_{i,n|n} ~=~	 -\mbox{inv}\big({\partial^2 \log f(z_{i,n}|\mathbf{y}_{ij,0:n},\theta)}\big/{\partial z^2_{i,n|n}}\big)$ &\\[0.2cm]\hline
	\end{tabular}
	\caption{Filtering algorithm on the interval $\{0,1,\ldots,N\}$. The algorithm takes as input the parameters $\theta$ and the data $\{\mathbf{y}_{ij,0:N}$, $\mathbf{u}_{ij,0:N}\}$ whereas returns as output the filtered states $\mathbf{z}_{i,0:N}$. Note that the notation $n|n-1$ indicates the prediction of the current $n$-th state given the previous $n-1$ whereas $n|n$ denotes the update of the predicted state.}
	\label{tab1}
\end{table}

Finally, to ensure that the unobserved sequence $\mathbf{z}_{i,0:N}$ is an approximate realization from $f(\mathbf{z}_{i,0:N}|\mathbf{y}_{ij,0:N})$ we need to refine the filtering results conditional on the whole observed data $\mathbf{y}_{ij,0:N}$. This task is achieved by means of a standard fixed-interval smoothing algorithm \cite{ansley1982geometrical,mendel1995lessons}, which uses the posterior filtering moments $z_{i,n|n}$ and $\lambda^2_{i,n|n}$ as input. The smoothing algorithm is described in Table \ref{tab2}.

\begin{table}[!h]
	\hspace*{-0cm}
	\begin{tabular}{llr}\\[0.05cm]\hline
		\small{\textit{Algorithm 2}} &{Fixed-interval smoothing algorithm}&\\[0.05cm]\hline
		$\mathbf{n=N:}$ & $z_{i,N|N} \sim \mathcall{N}(z_{i,N|n},\lambda^2_{i,N|n})$ & \textsc{initialization}\\[0.3cm]
		& $\lambda^2_{i,N|N} = 1$ & \\[0.3cm]
		$\mathbf{n<N:}$ & $z_{i,n|N} ~=~ z_{i,n|n} + \biggl(\frac{\lambda^2_{i,n|n}}{\lambda^2_{i,n+1|n}}\biggr)(z_{i,n+1|N}-z_{i,n+1|n})$ & \textsc{backward update}\\[0.3cm]
		& $\lambda^2_{i,n|N} ~=~ \lambda^2_{i,n|n} + \biggl(\frac{\lambda^2_{i,n|n}}{\lambda^2_{i,n+1|n}}\biggr)^2(\lambda^2_{i,n+1|N}-\lambda^2_{i,n+1|n})$ & \\[0.3cm]\hline
	\end{tabular}
	\caption{Backward smoothing algorithm over the interval $\{N,N-1,\ldots,0\}$. The algorithm takes as input the filtering solutions whereas returns as output the smoothed states $\mathbf{z}_{i,0:N}$ conditional on the whole set of data $\mathbf{y}_{ij,0:N}$. Note that the the notation $.|n$ refers to the filtering solutions whereas $.|N$ indicates the smoothing results.} 
	\label{tab2}
\end{table}

\section*{Appendix B: Posteriors computation and estimation}

\noindent In what follows, we describe the steps for computing the term $f(\boldsymbol{\Theta}|\mathbf{Y})$. First, we note that the array $\boldsymbol\Theta$ consists of two blocks of parameters associated to the observation equation and the stimuli equation of the model \eqref{eq1_2}-\eqref{eq1_4}, namely $J$ scalars $\{\beta_1,\ldots,\beta_J\}$ paired with the set of stimuli/trials and two parameters $\{\kappa_1,\kappa_2\}$ for the vonMises concentrations. To simplify the computations in the Metropolis-Hastings algorithm, the terms $\{\kappa_1,\kappa_2\}$ can be distinctively determined prior running the MH algorithm \cite{marin2005bayesian}. Moreover, since we are not interested in the posterior distributions of these parameters, as long as they are not involved in the state-space dynamics, we compute them using the following Maximum-Likelihood estimators \cite{banerjee2005clustering}:    
\begin{align}
	& \hat{\kappa}_1 = I^{-1}\biggl(\frac{\sum_{i=1}^I \sum_{j=1}^J \sum_{n=0}^N u_{ijn} \cos\bigl(y_{ijn}-\mu_1\bigr)}{\sum_{i=1}^I \sum_{j=1}^J \sum_{n=0}^N u_{ijn}}\biggl) \label{11_1}\\[0.3cm]
	& \hat{\kappa}_2 = I^{-1}\biggl(\frac{\sum_{i=1}^I \sum_{j=1}^J \sum_{n=0}^N (1-u_{ijn}) \cos\bigl(y_{ijn}-\mu_2\bigr)}{\sum_{i=1}^I \sum_{j=1}^J \sum_{n=0}^N (1-u_{ijn})}\biggl) \label{11_2}	
\end{align}
where $\mu_1$ and $\mu_2$ are the location parameters fixed by the experimenter, $u_{ijn}$ is defined as in Eq. \eqref{eq2_2b}, whereas $I^{-1}$ is the inverse of the modified Bessel function which is evaluated numerically \cite{abramowitz1972handbook}. 
Given the above results and the constrains $\boldsymbol{\sigma}_{I\times 1} = \mathbf 1_I$, the array of parameters simply reduces to $\boldsymbol{\theta}_{J\times 1}$. This simplifies the inner-working of the MH algorithm as it now works on a smaller and more compact parameter space. To proceed further, the decomposition \eqref{eq4_1} involves the following definition for the MH proposal density \cite{andrieu2010particle}:
\begin{equation}\label{eq12_1}
	q(\{\boldsymbol{\theta}^{(t)},\mathbf{Z}^{(t)}\}|\{\boldsymbol{\theta}^{(t-1)},\mathbf{Z}^{(t-1)}\}) ~=~ q(\boldsymbol{\theta}^{(t)}|\boldsymbol{\theta}^{(t-1)})f(\mathbf{Z}^{(t)}|\mathbf{Y})
\end{equation}
\noindent where $f(\mathbf{Z}^{(t)}|\mathbf{Y})$ is evaluated through filtering/smoothing. This is appealing since the posterior density $f(\mathbf{Z},\boldsymbol{\Theta}|\mathbf{Y})$ from which it might be difficult to sample from, reduces now to $f(\boldsymbol{\Theta}|\mathbf{Y})$ that is conveniently defined on a smaller parameters space \cite{andrieu2009pseudo}. In our case, we can set: 
\begin{equation}\label{eq12_2}
	q(\boldsymbol\theta^{(t)}|\boldsymbol\theta^{(t-1)}) ~=~\mathcall{N}(\boldsymbol\theta^{(t-1)},\boldsymbol\Sigma^{(t)})	
\end{equation}
with $\boldsymbol\Sigma^{(t)}$ being a suitable $J\times J$ covariance matrix. Consequently, the MH acceptance ratio is as follows:
\begin{equation}\label{eq12_3}
	\alpha^{(t)} ~=~ \frac{f(\mathbf{Y}|\boldsymbol\theta^{(t)}) ~q(\boldsymbol\theta^{(t-1)}|\boldsymbol\theta^{(t)}) ~f(\boldsymbol\theta^{(t)})}{f(\mathbf{Y}|\boldsymbol\theta^{(t-1)}) ~q(\boldsymbol\theta^{(t)}|\boldsymbol\theta^{(t-1)}) ~f(\boldsymbol\theta^{(t-1)})}
\end{equation}
\noindent where $f(\mathbf{Y}|\boldsymbol{\theta})$ is the density for the marginal likelihood computation, $f(\boldsymbol\theta)$ indicates the prior density over the parameters, whereas $q(\boldsymbol\theta^{(t)}|\boldsymbol\theta^{(t-1)})$ is the MH proposal density. Note that under \eqref{eq12_2}, the terms $q(\boldsymbol\theta^{(t)}|\boldsymbol\theta^{(t-1)})$ and $q(\boldsymbol\theta^{(t-1)}|\boldsymbol\theta^{(t)})$ in Eq. \eqref{eq12_3} vanish as they refer to the same probability value. This yields to: 
\begin{equation}\label{eq12_4}
	\alpha^{(t)} ~=~ \frac{f(\mathbf{Y}|\boldsymbol\theta^{(t)})  ~f(\boldsymbol\theta^{(t)})}{f(\mathbf{Y}|\boldsymbol\theta^{(t-1)}) ~f(\boldsymbol\theta^{(t-1)})}\vspace{0.3cm}
\end{equation}
where the MH ratio is now expressed as a function of the marginal likelihood and the priors. The term $f(\mathbf{Y}|\boldsymbol\theta)$ can be easily computed as a byproduct of the filtering calculations (see Appendix C). 
Finally, the choice of a well-suited covariance matrix $\boldsymbol\Sigma^{(t)}$ for the proposal distribution is crucial in order to achieve chains' convergences. In our context, we used the Haario's adaptive solution where the proposal covariance is iteratively adapted during the chains using the current proposal covariance up to the adaptation step \cite{haario2001adaptive}. The complete MH algorithm is summarized in Table \ref{tab3}.

\begin{table}[!h]
	\hspace*{-0.0cm}
	\begin{tabular}{llr}\\[0.05cm]\hline
		\small{\textit{Algorithm 3}} &{Metropolis-Hasting algorithm}&\\[0.05cm]\hline
		$\mathbf{t=0:}$ & \small{Set} ~\normalsize$\boldsymbol\theta^{(t)} = \boldsymbol\theta^{(0)}$, $\boldsymbol\Sigma^{(t)} = \boldsymbol\Sigma^{(0)}$ & \textsc{initialization}\\
		& \small{Run Algorithms 1-2 to get} ~\normalsize$\mathbf{Z}^{(0)}\sim f\bigl(\mathbf{Z}|\mathbf{Y},\boldsymbol{\theta^{(0)}}\bigr)$ & \\
		& \small{Compute} ~\normalsize$f(\mathbf{Y}|\boldsymbol{\theta^{(0)}})~$ &\\[0.3cm] 
		$\mathbf{t>0:}$ & $\boldsymbol{\theta}^{\star} ~\sim~ \mathcall{N}(\boldsymbol{\theta}^{(t-1)},\boldsymbol\Sigma^{(t)})$ & \textsc{m-h loop}\\
		& \small{Run Algorithms 1-2 to get} ~\normalsize$\mathbf{Z}^\star\sim f\bigl(\mathbf{Z}|\mathbf{Y},\boldsymbol{\theta^{\star}}\bigr)$ & \\
		& \small{Compute} ~\normalsize$f(\mathbf{Y}|\boldsymbol{\theta^{\star}})~$ &\\ 
		& \small{Compute} ~\normalsize$\alpha^{(t)}$ \small{from Eq. \eqref{eq12_4}}\normalsize &\\[0.3cm]
		& \small{Get }~\normalsize $r \sim \mathcall{U}(0,1)$ & \textsc{accept/reject}\\
		& \small{Set} ~\normalsize$\boldsymbol\theta^{(t)}=\boldsymbol\theta^\star$, $\mathbf{Z}^{(t)}=\mathbf{Z}^\star$ ~\small{ if }\normalsize $\alpha^{(t)} \leq r$ & \\	
		& \small{Set} ~\normalsize$\boldsymbol\theta^{(t)}=\boldsymbol\theta^{(t-1)}$, $\mathbf{Z}^{(t)}=\mathbf{Z}^{(t-1)}$ ~\small{ if }\normalsize $\alpha^{(t)} > r$ &\textsc{}\\[0.3cm]
		& \small{Run} ~\normalsize$\boldsymbol\Sigma^{(t+1)} \leftarrow \mbox{adapt}\bigl(\boldsymbol\Sigma^{(t)}\bigr)$ ~\small{ see \cite{haario2001adaptive}} &\textsc{adapting phase}\\\hline
	\end{tabular}
	\caption{Marginal Metropolis-Hastings to estimate $\boldsymbol\beta_{J\times 1}$. Note that $\mathcall{U}(0,1)$ indicates the Uniform distribution over the interval $[0,1]$ whereas the adaptive phase can be performed at each iteration $t$ of the chain or rather after a fixed interval $t+H$ (with $H>1$).}
	\label{tab3}
\end{table}

\noindent 
\section*{Appendix C: Marginal Likelihood computation}  

\noindent The marginal likelihood $f(\mathbf Y|\boldsymbol{\theta})$ is computed as follows. Let the observed-data log likelihood be:
\begin{align}\label{eq9}
	\log \mathcall{L}(\theta|\mathbf{Y}) &=\sum_{i=1}^I\sum_{j=1}^J   \log f(\mathbf{y}_{ij,0:n}|\theta)\nonumber\\
	&=\sum_{i=1}^I\sum_{j=1}^J \Biggl( \sum_{n=0}^N \log f(y_{ijn}|\mathbf{y}_{ij,0:n-1},\theta) \Biggr)\\
	&=\sum_{i=1}^I\sum_{j=1}^J \Biggl(\sum_{n=0}^N \log \int_{\mathbb R} f(y_{ijn}|z_{i,n},\theta) f(z_{i,n}|\mathbf{y}_{ij,0:n-1},\theta) ~dz_{i,n} \Biggr)\nonumber
\end{align}
According to the standard prediction error decomposition, this functional has been factorized as a function of both the measurement model density \eqref{eq2_2} and the one-step ahead predictive density \eqref{eq6_4} \cite{de1988likelihood}. By substituting the above definitions with our model's densities, we get the following marginal likelihood:

\begin{align}\label{eq10}
	\log \mathcall{L}(\theta|\mathbf{Y}) &= \sum_{i=1}^I\sum_{j=1}^J \sum_{n=0}^N \log \Biggl(\bigintssss_{\mathbb{R}} ~\bigl(\pi_{ijn} \exp(\cos(y_{ijn}-\mu_1)^{\kappa_1}\bigr)^{u_{ijn}} \cdot\nonumber\\
	&\cdot \bigl((1-\pi_{ijn}) \exp(\cos(y_{ijn}-\mu_2)^{\kappa_2}\bigr)^{(1-u_{ijn})} \cdot \nonumber \\
	&\cdot\exp\biggl(-\frac{(z_{i,n}-z_{i,n|n-1})^2}{2(\sigma^2_i +\lambda^2_{i,n-1|n-1})}\biggr) ~dz_{i,n} \Biggr)
\end{align} 

\noindent where the von-Mises and Gaussian densities have been written by dropping the constant terms. Using the predictive moments $z_{i,n|n-1}$ and $\lambda^2_{i,n|n-1}$ from the filter algorithm (see Table \ref{tab1}), a recursion on $\{0,1,\ldots,N\}$ can be written for the likelihood computation, which consists of solving $N$ integrals over the support of the r.v. $Z_{i,n}$. Since no analytical solutions are available for the functional above, numerical integration methods can be used to solve the integrals \cite{shampine2008vectorized}.

\clearpage
\bibliographystyle{plain}
\bibliography{biblio}

\end{document}


\maketitle

\vspace{2.5cm}
\begin{center}
	\begin{minipage}{14cm}
		\rule{14cm}{0.01cm}	
		\tableofcontents
		\rule{14cm}{0.01cm}
	\end{minipage}
\end{center}

\clearpage
\section{Evaluating the state-space model}
\subsection{Computation time}

\begin{figure}
	\hspace*{-1cm}
	\includegraphics[scale=0.45,angle=0]{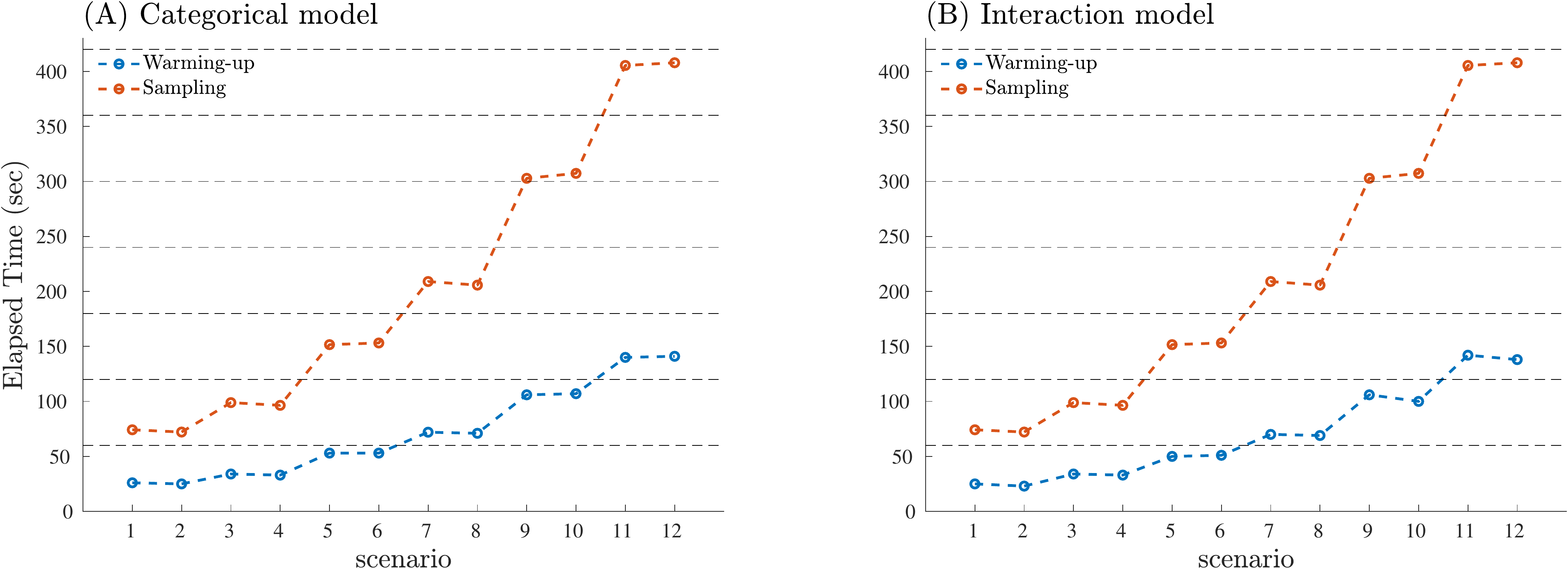}
	\caption{Computation time: Elapsed time (sec.) with regards to warming-up (blue line) and sampling (red line) for the simplest categorical (A) and interaction (B) models. Note that dotted gray lines indicate minutes.}
	\label{fig1}
\end{figure}

In this section we evaluate the time required by the algorithm included in Table A.3 of the manuscript to completely fit the state-space model. The algorithm has been implemented in Matlab R2018a  and it is freely available online at \url{https://github.com/antcalcagni/SSM_mousetracking}. We tested two instances of our model, namely the simple model with a categorical variable  and the complete model including categorical and continuous variables together with their interaction. The simulation study was obtained by varying the type of model (\textit{categorical}, \textit{interaction}), the number of statistical units $I = (12,25,50)$, the number of trials $J=(12,28)$ and the number of levels of the categorical variable $K=(2,4)$. The time step was kept fixed $N=61$. The study involved $I\times J\times K = 12$ scenario for two type of models (see Table \ref{tab1}). For each scenario, data and model structure were kept fixed, one single chain was used without parallelization, warming-up and sampling periods were equal to 2000 samples (warm-up = 1000 samples), and the adaptive parameter of the MH-loop was equal to $H=25$. The elapsed time was measured in seconds using the standard built-in Matlab function. Computations were performed on Intel Core i7-7500U 4-core 2.70GHz, Ubuntu 18.04 64-bit, 8GB Ram. Figure \ref{fig1} shows the results for both types of models. As expected, the elapsed time increased exponentially as a function of the complexity of the scenario. In both models, the most simple scenario (scenario no. 1) required less then one minute to complete the warming-up and about one minute to complete the sampling period. Instead, the medium scenario (scenario no. 6) required about one minute to complete the warming-up and less then three minutes to complete the sampling phase. Finally, the most complex scenario (scenario no. 12) required less than three minutes to finalize the warming-up and about seven minutes to complete the sampling period.  

\begin{table}[!h]
	\centering
	\begin{tabular}{lc}
		scenario & parameters \\\hline\hline
		1 & I=12,J=12,K=2\\
		2 & I=12,J=12,K=4\\
		3 & I=12,J=28,K=2\\
		4 & I=12,J=28,K=4\\
		5 & I=25,J=12,K=2\\
		6 & I=25,J=12,K=4\\
		7 & I=25,J=28,K=2\\
		8 & I=25,J=28,K=4\\
		9 & I=50,J=12,K=2\\
		10 & I=50,J=12,K=4\\
		11 & I=50,J=28,K=2\\
		12 & I=50,J=28,K=4\\\hline\hline
	\end{tabular}
	\caption{Computation time: Simulated scenario for both categorical and interaction models}
	\label{tab1}
\end{table}

\subsection{Simulation study}

In this section, we report the results of a simulation study to evaluate the ability of the model to (i) \textit{recover} the true parameters $\boldsymbol{\beta}$ and (ii) \textit{resemble} the observed data $\mathbf{Y}$. We designed the simulation under the Bayesian framework of statistical modeling. In particular, four factors were varied in a complete factorial design with 16 scenarios: type of model = \{categorical, interaction\}, $I=\{12,25\}$, $J=\{12,28\}$, $K=\{2,4\}$. The combinations $\text{type of model}\times I \times J \times K$ were evaluated $M=5000$ times by getting a total of 40000 final samples. The levels of $I$, $J$, and $K$ were chosen to represent the number of individuals, trials, and variables commonly encountered in experimental psychology. The simulation was performed on a (remote) HPC machine based on 16 cpu Intel Xeon CPU E5-2630L v3 1.80GHz, 16x4GB Ram. 

\subsubsection{Simulation design}
\noindent The simulation workflow proceeded as follows:
\begin{enumerate}
	\item The hyperparameters $\boldsymbol\mu_\beta$ and $\boldsymbol\Sigma_\beta$ of the prior $f(\boldsymbol{\beta}) = \mathcall N(\boldsymbol\mu_\beta,\boldsymbol\Sigma_\beta) $ were kept fixed, $\mu_1=2.75$, $\mu_2=0.75$, $\kappa_1=\kappa_2=100$, $X\sim \mathcall U_d(-3,3)$, $\mathbf D=\mathbf{I}_{K\times 1} \otimes \mathbf{1}_{\frac{J}{K}\times 1}$ (with $\otimes$ being the Kronecker product)\\
	\item For each scenario $s=1,\ldots,16$, the parameters were sampled from the priors $\boldsymbol{\beta}_{1,\ldots,M}\sim f(\boldsymbol{\beta})$ and a series of $M$ datasets $\mathbf{Y}^s_1,\ldots,\mathbf{Y}^s_M$ were generated according to the model equations (note that the generic dataset is of the form $\mathbf Y^s_{I\times J\times N}$)\\
	\item For each of the datasets $\mathbf{Y}^s_1,\ldots,\mathbf{Y}^s_M$, the model was fitted by obtaining the array of latent states and parameter posteriors: $$\mathcall R = \{(\mathbf Z_1^s,\ldots,\mathbf Z^s_M), (f(\boldsymbol\beta|\mathbf{Y})_1^s,\ldots,f(\boldsymbol\beta|\mathbf{Y})_M^s)\}$$ where the generic array $\mathbf Z^s$ is of the form $\mathbf Z^s_{I\times N}$. The model was run using 10000 (samples) x 8 (chains), with warming-up of 2500 samples. Matlab Parallel computing was used to run the chains. 
\end{enumerate}

\subsubsection{Outcome measures}
\noindent The \textit{recovery problem} was evaluated for $s=1,\ldots,16$ by means of the overlapping measure between symmetric distributions:
$$
\Lambda\Big(f(\boldsymbol{\beta})^s,f(\boldsymbol{\beta}|\mathbf Y)^s_{1,\ldots,M}\Big) = \frac{1}{M} \sum_{m=1}^{M}\int_{\mathbb R^n} \min\Big( f(\boldsymbol{\beta})^s,f(\boldsymbol{\beta}|\mathbf Y)^s_{m} \Big)
$$
which measures the amount of agreement between two distributions, with $\Lambda=0$ indicating no overlapping. In our context, as $f(\boldsymbol{\beta})$ and $f(\boldsymbol{\beta}|\mathbf{Y})$ are symmetric with a unique mode, we interpret $\Lambda$ as density similarity where $\Lambda=1$ means that the two distributions are the same\footnote{Schmid, F., \& Schmidt, A. (2006). Nonparametric estimation of the coefficient of overlapping—theory and empirical application. \textit{Computational statistics \& data analysis}, 50(6), 1583-1596.}. \\

\noindent The \textit{resembling problem} was instead assessed by means of the $\mathbf{Y}|\mathbf{\tilde{Y}}$-discrepancy measure: 
$$
\overline{\text{AoR}}^s = \frac{1}{M} \sum_{m=1}^M\text{AoR}\Big(\mathbf{Y}^s_m,\mathbf{\tilde{Y}}^s_m\Big)
$$
where:
$$\text{AoR}\Big(\mathbf{Y},\mathbf{\tilde{Y}}\Big) = 100\cdot \frac{\min\big(|| \mathbf{y}^\dagger ||, || \text{vec}(\mathbf{Y}) || \big)}{\max\big(|| \mathbf{y}^\dagger ||, || \text{vec}(\mathbf{Y}) || \big)} \quad\text{with}\quad \mathbf{y}^\dagger = \frac{\text{vec}(\mathbf{\tilde{Y}})^T\text{vec}(\mathbf{Y})}{||\text{vec}(\mathbf{Y})||} \cdot \frac{\text{vec}(\mathbf{Y})}{||\text{vec}(\mathbf{Y})||}$$
which measures the \textit{amount of reconstruction}\footnote{Sadtler, P. T., Quick, K. M., et al. (2014). Neural constraints on learning. \textit{Nature}, 512(7515), 423-426.} of the observed array $\mathbf{Y}$ via the reproduced array $\mathbf{\tilde{Y}}$. The measure AoR takes values in [0,100] with 0\% indicating that no reconstruction occurred at all.

\subsubsection{Results}

The results of the simulation study are described in Tables \ref{tab2a}-\ref{tab3a}. Overall, both the models got acceptable fit with regard to recover the true model structure ($\Lambda$ index) and the adequacy in reproducing the input data (AoR index). As expected, the goodness-of-fit increased as a function of the number of trials $J$ and the number of individuals $I$. 

\begin{table}[!h]
	\centering
	\begin{tabular}{lccccccc}
		&&& $\text{AoR}$ & {${\Lambda}_{\gamma_1}$}& {${\Lambda}_{\gamma_2}$}& {${\Lambda}_{\gamma_3}$}& {${\Lambda}_{\gamma_4}$}\\\hline\hline
		\multirow{4}{*}{\tiny{$I=12$}} & \tiny{$J=12,K=2$} &&0.688&0.816&0.790&&\\
		&\tiny{$J=12,K=4$} &&0.746&0.825&0.862&0.853&0.794\\
		&\tiny{$J=28,K=2$} &&0.732&0.828&0.848&&\\
		&\tiny{$J=28,K=4$} &&0.799&0.874&0.807&0.830&0.831\\\hline
		\multirow{4}{*}{\tiny{$I=25$}} & \tiny{$J=12,K=2$} && 0.780&0.848&0.796&&\\
		&\tiny{$J=12,K=4$} &&0.808&0.749&0.841&0.805&0.895\\
		&\tiny{$J=28,K=2$} &&0.753&0.809&0.844&&\\
		&\tiny{$J=28,K=4$} &&0.811&0.828&0.866&0.833&0.857\\\hline\hline		
	\end{tabular}
	\caption{Simulation study (Categorical model): AoR and $\Lambda$ indices for each level of the simulation design.}
	\label{tab2a}	
\end{table}

\begin{table}[!h]
	\centering
	\begin{tabular}{lccccccccccc}
		&&& $\text{AoR}$ & {${\Lambda}_{\gamma_1}$}& {${\Lambda}_{\gamma_2}$}& {${\Lambda}_{\gamma_3}$}& ${\Lambda}_{\gamma_4}$ & ${\Lambda}_{\eta}$ & ${\Lambda}_{\delta_1}$ & {${\Lambda}_{\delta_2}$} & {${\Lambda}_{\delta_3}$}\\\hline\hline
		\multirow{4}{*}{\tiny{$I=12$}} & \tiny{$J=12,K=2$} && 0.695 & 0.809 & 0.843 &  &  & 0.637 & 0.841 &  &  \\
		&\tiny{$J=12,K=4$} && 0.720 & 0.909 & 0.880 & 0.943 & 0.754 & 0.685 & 0.794 & 0.872 & 0.827\\
		&\tiny{$J=28,K=2$} &&0.807 & 0.813 & 0.831 &  &  & 0.684 & 0.835 &  & \\
		&\tiny{$J=28,K=4$} &&0.710 & 0.955 & 0.946 & 0.863 & 0.902 & 0.674 & 0.933 & 0.888 & 0.900\\\hline
		\multirow{4}{*}{\tiny{$I=25$}} & \tiny{$J=12,K=2$} && 0.751 & 0.816 & 0.843 &  &  & 0.737 & 0.789 & & \\
		&\tiny{$J=12,K=4$} &&0.861 & 0.919 & 0.893 & 0.885 & 0.910 & 0.675 & 0.838 & 0.897 & 0.835\\
		&\tiny{$J=28,K=2$} &&0.794 & 0.862 & 0.842 &  &  & 0.701 & 0.762 &  & \\
		&\tiny{$J=28,K=4$} &&0.873 & 0.875 & 0.848 & 0.981 & 0.806 & 0.726 & 0.887 & 0.870 & 0.869\\\hline\hline
	\end{tabular}
	\caption{Simulation study (Interaction model): AoR and $\Lambda$ indices for each level of the simulation design.}
	\label{tab3a}	
\end{table}

\section{Case study}
\subsection{Convergences of the MCMC chains}

In this section we illustrate the convergence diagnostics for the three case studies described in Section 4 of the manuscript.

\begin{figure}[!h]
	\includegraphics[scale=0.38,angle=0]{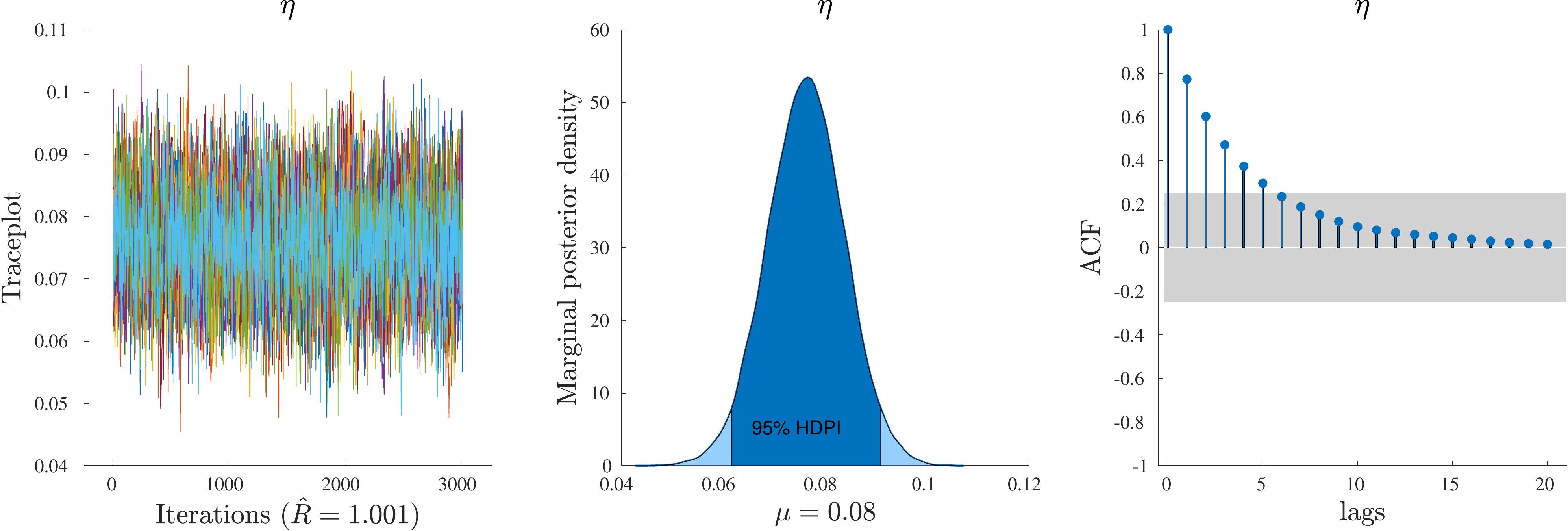}
	\caption{Case study (Model 2): MCMC diagnostics. Left panel: Traceplot of the chains in different colors and Gelman-Rubin $\hat R$ index. Middle panel: Marginal posterior density with posterior means $\mu$ and 95\% HDPI. Right panel: Autocorrelation function (ACF) for the collapsed chains.}
	\label{fig_mod_3}
\end{figure}

\begin{figure}[!h]
	\hspace*{-1.5cm}
	\includegraphics[scale=0.4,angle=0]{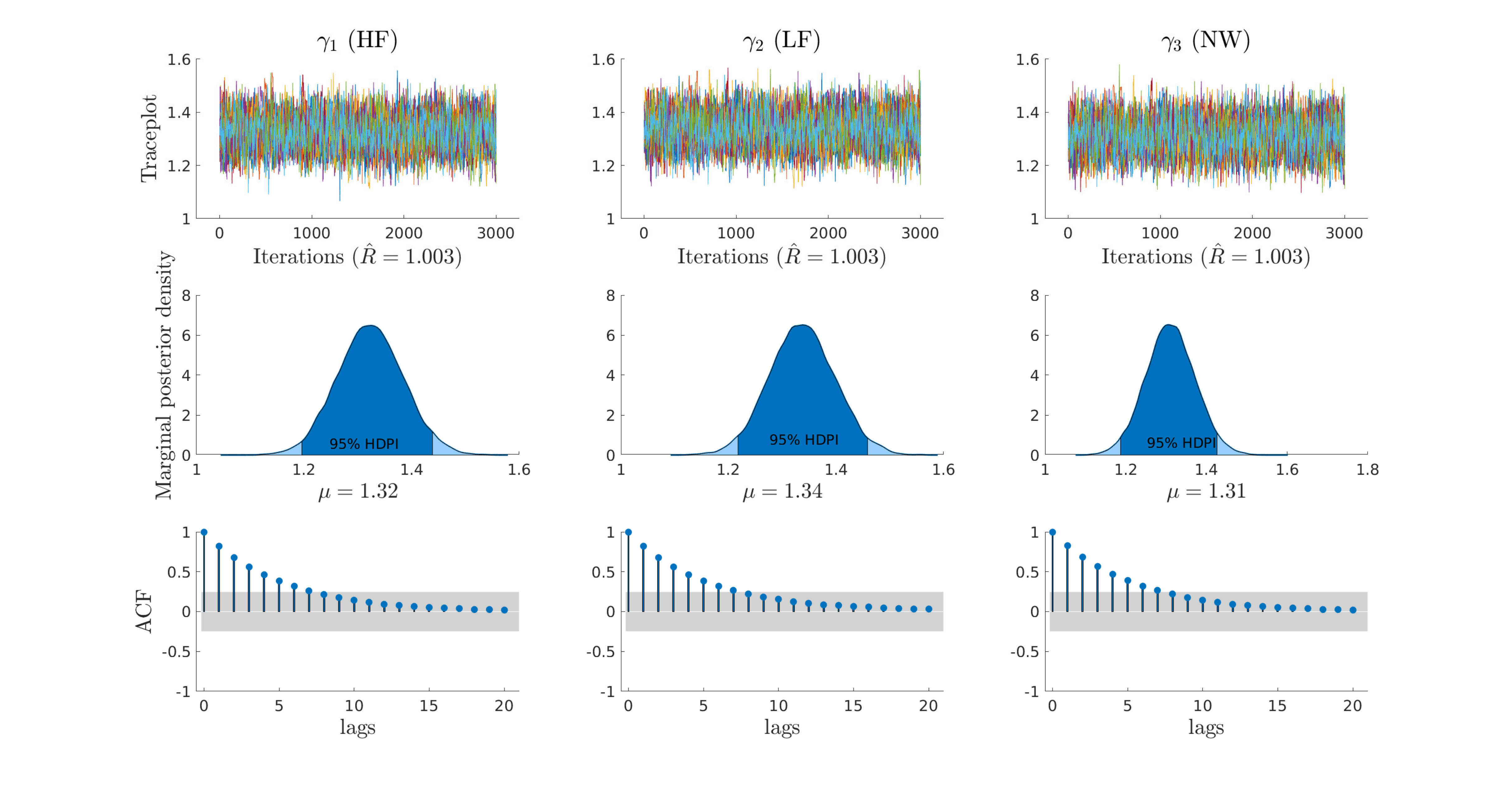}
	\caption{Case study (Model 1): MCMC diagnostics. First row: Traceplot of the chains in different colors and Gelman-Rubin $\hat R$ index. Second row: Marginal posterior densities with posterior means $\mu$ and 95\% HDPI. Third row: Autocorrelation function (ACFs) for the collapsed chains.}
	\label{fig_mod_1}
\end{figure}

\begin{figure}[!h]
	\hspace*{2cm}
	\includegraphics[scale=0.5,angle=90]{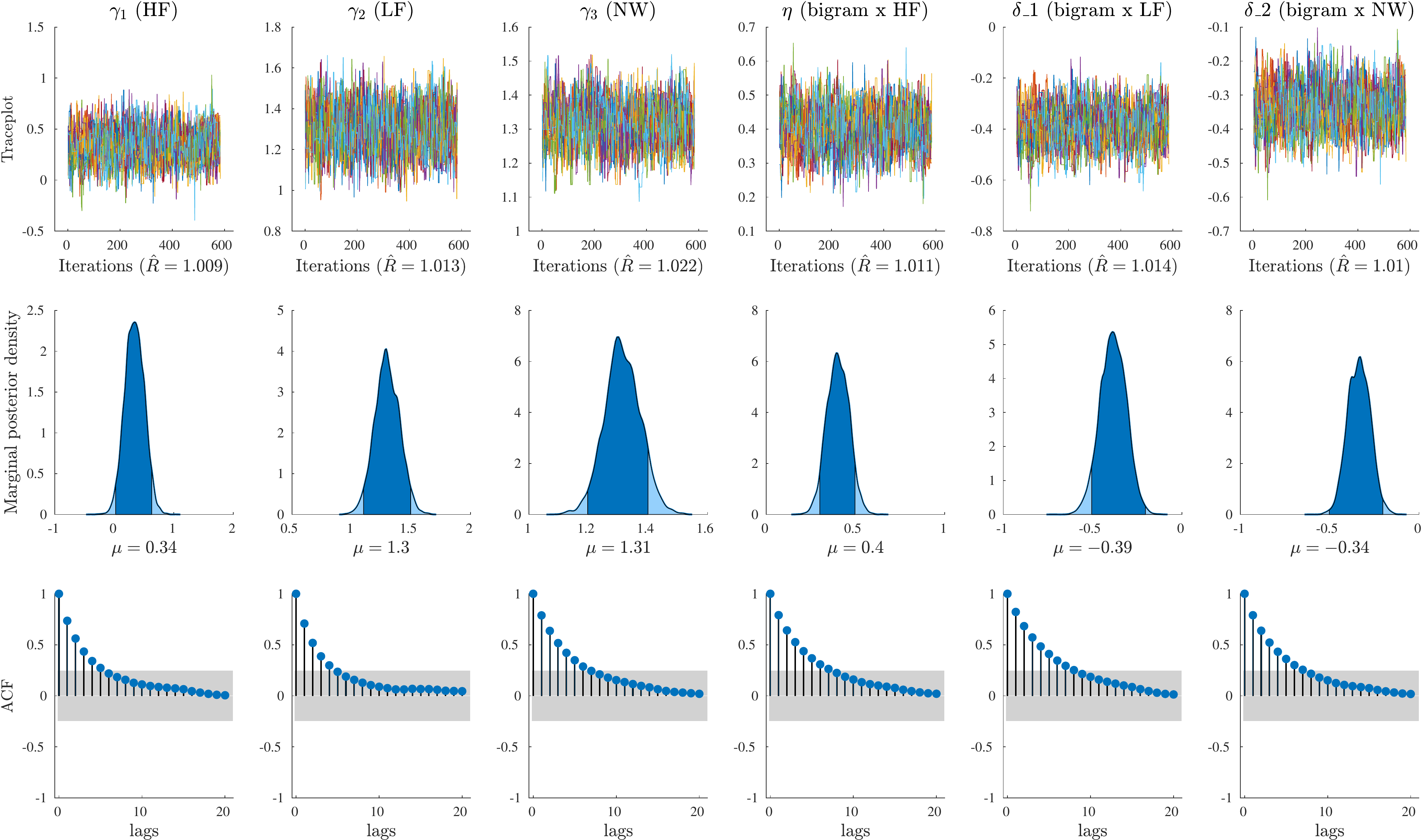}
	\caption{Case study (Model 3): MCMC diagnostics. First row: Traceplot of the chains in different colors and Gelman-Rubin $\hat R$ index. Second row: Marginal posterior densities with posterior means $\mu$ and 95\% HDPI. Third row: Autocorrelation function (ACFs) for the collapsed chains.}
	\label{fig_mod_5}
\end{figure}

\clearpage
\subsection{Fit of the model}

The adequacy of the state-space model to reproduce the observed data was computed by means of a simulation-based approach. In particular, given the posteriors of the parameters $\hat{\boldsymbol{\beta}}$ and filtered/smoothed latent states $\hat{\mathbf{Z}}$, $M$ new (simulated) datasets $(\mathbf{Y}^*_1,\ldots,\mathbf{Y}^*_M)$ were generated according to the estimated model structure. For each new dataset, the AoR discrepancy measure was computed:
$$\text{AoR} = 100\cdot \frac{\min\big(|| \mathbf{y}^\dagger ||, || \text{vec}(\mathbf{Y}) || \big)}{\max\big(|| \mathbf{y}^\dagger ||, || \text{vec}(\mathbf{Y}) || \big)} \quad\text{with}\quad \mathbf{y}^\dagger = \frac{\text{vec}(\mathbf{Y}^*_m)^T\text{vec}(\mathbf{Y})}{||\text{vec}(\mathbf{Y})||} \cdot \frac{\text{vec}(\mathbf{Y})}{||\text{vec}(\mathbf{Y})||} \quad\quad m=1,\ldots,M$$
In addition, to further evaluate the amount of reconstruction, we computed two kind of AoR measures:
\begin{itemize}
	\item \textbf{Overall AoR:} percentage of data reconstruction based on the overall $I\times J\times N$ observed matrix $\mathbf Y$.\\[0.1cm]
	\item \textbf{By-subject AoR:} percentage of data reconstruction based on the $J\times N$ observed matrix $\mathbf Y_i$ for each subject $i=1,\ldots, I$. The index allows for evaluating the adequacy of the model to reconstruct the individual-based set of trajectories.
\end{itemize}

\begin{figure}[!h]
	\hspace*{-2.5cm}
	\includegraphics[scale=0.5,angle=0]{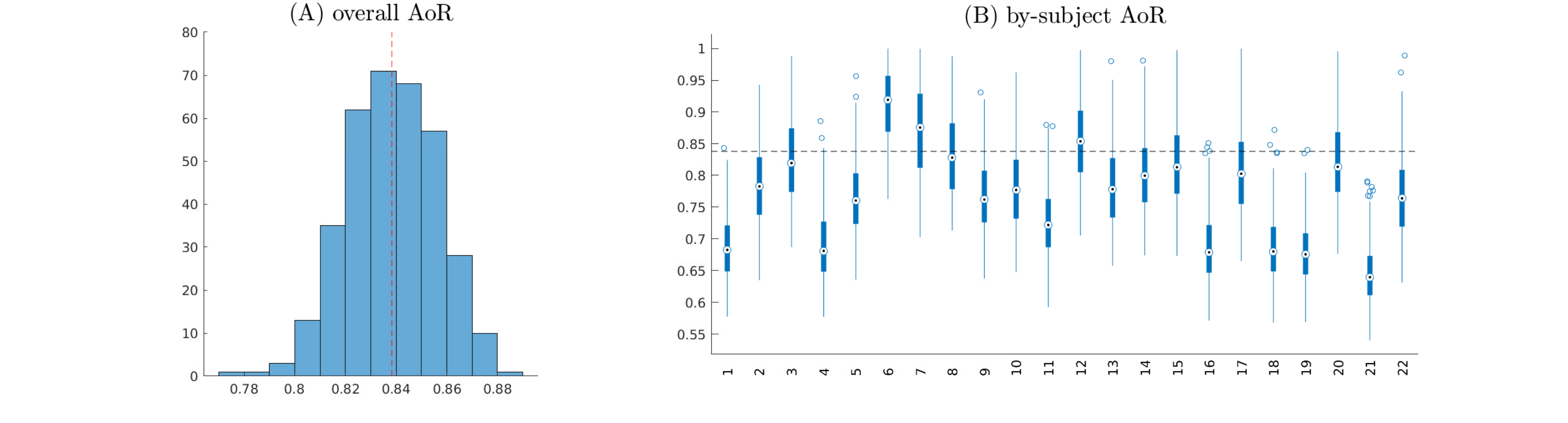}
	\caption{Goodness of fit (Model 1): (A) Overall amount of reconstruction and (B) By-subject amount of reconstruction. Note that dotted lines represent the mean of the overall AoR index.}
	\label{fig2}
\end{figure}

\begin{figure}[!h]
	\hspace*{-2.5cm}
	\includegraphics[scale=0.5,angle=0]{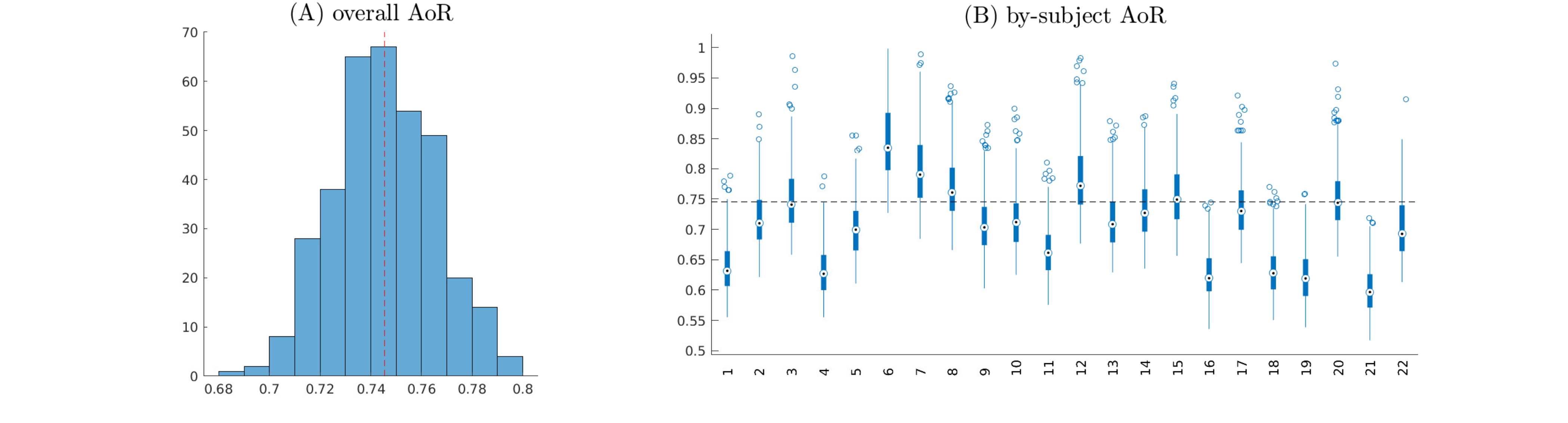}
	\caption{Goodness of fit (Model 2): (A) Overall amount of reconstruction and (B) By-subject amount of reconstruction. Note that dotted lines represent the mean of the overall AoR index.}
	\label{fig3}
\end{figure}

\begin{figure}[!h]
	\hspace*{-2.5cm}
	\includegraphics[scale=0.5,angle=0]{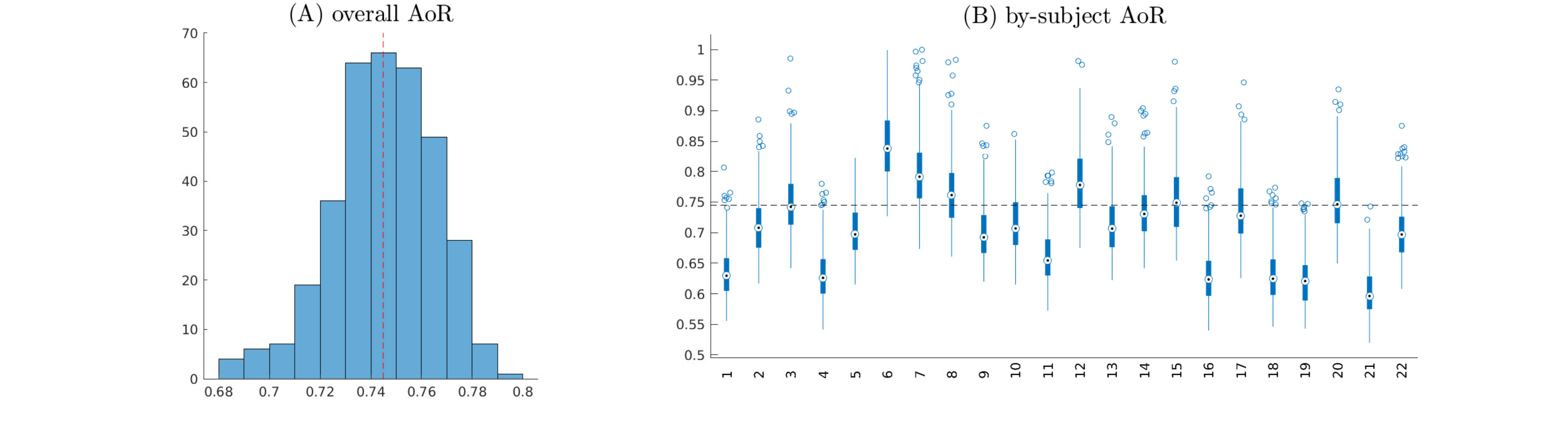}
	\caption{Goodness of fit (Model 3): (A) Overall amount of reconstruction and (B) By-subject amount of reconstruction. Note that dotted lines represent the mean of the overall AoR index.}
	\label{fig4}
\end{figure}

\begin{figure}[!h]
	\hspace*{1.5cm}
	\includegraphics[scale=0.8,angle=90]{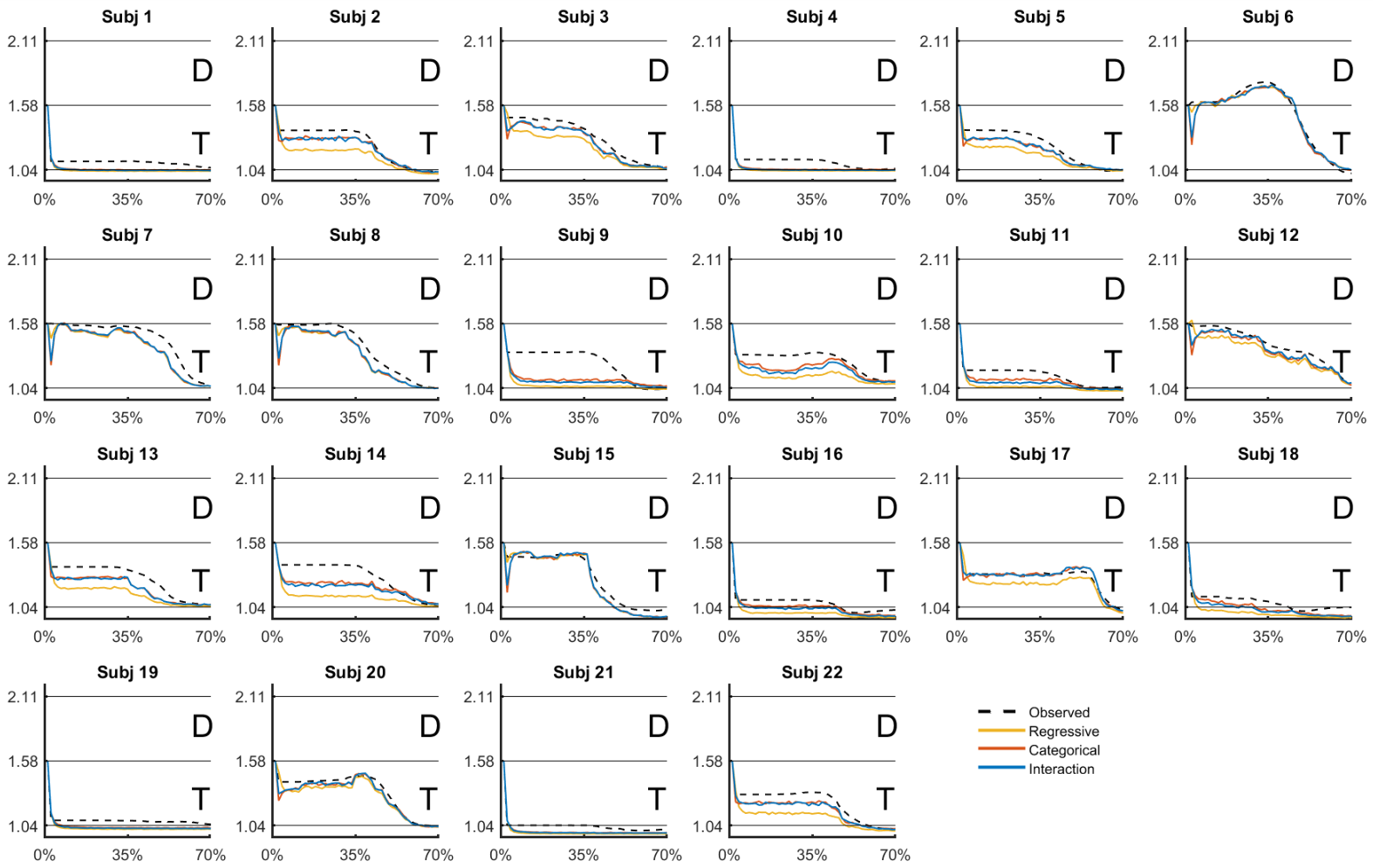}
	\caption{Goodness of fit: Observed and predicted individual (mean) trajectories for each of three models. Note that dotted lines represent observed trajectories whereas the upper and lower halves indicate distractor (D) and target (T) hemispaces, respectively. }
	\label{fig5}
\end{figure}

The simulation was performed on a (remote) HPC machine based on 16 cpu Intel Xeon CPU E5-2630L v3 1.80GHz, 16x4GB Ram. Figures \ref{fig2}-\ref{fig4} show the results of AoR indices for the three models. Figure \ref{fig5} shows the observed mean trajectories for each individual $\mathbf Y_i$, $i=1,\ldots,I$ against the predicted trajectories by the three models. 

